\documentclass[aps,amsmath,amsfonts,floatfix,twocolumn,superscriptaddress]{revtex4-1}
\usepackage{graphicx}
\usepackage[colorlinks,linkcolor=blue,citecolor=blue,urlcolor=blue]{hyperref}
\usepackage{natbib}

\usepackage{ulem}

\begin{document}
%\title{Interplay of coherence and interaction in light propagation through waveguide QED lattices}
\title{A modified quasi-classical analysis to capture the effects of strong interaction in open QED lattices}
\date{\today}
\author{Tarush Tiwari}
\affiliation{School of Biochemical Engineering, Indian Institute of Technology (Banaras Hindu University), Varanasi 221005, India.}
\affiliation{Department of Physics and Applied Physics, University of Massachusetts, Lowell, Massachusetts 01854, USA.}
\author{Kuldeep K Shrivastava}
\affiliation{Department of Physics, Indian Institute of Technology (Banaras Hindu University), Varanasi 221005, India.}
\author{Dibyendu Roy}
\affiliation{Raman Research Institute, Bangalore 560080, India.}
\author{Rajeev Singh}
\affiliation{Department of Physics, Indian Institute of Technology (Banaras Hindu University), Varanasi 221005, India.}
\begin{abstract}
We investigate the role of optical nonlinearity in light propagation through two different one-dimensional open QED lattices, namely a chain of qubits with direct coupling between the nearest neighbors and a chain of connected resonators to each of which a qubit is side-coupled. Using the more accurate truncated Heisenberg-Langevin equations method we show a reduction of light transmission with increasing intensity in these lattices due to effective photon-photon interactions and related photon blockade mediated by nonlinearity in qubits. In contrast to the direct-coupled qubits, we find a revival in the light transmission in the side-coupled qubits at relatively higher intensities due to saturation of qubits by photons. We find that in absence of bulk dissipation the standard quasi-classical analysis fails to capture the reduction in light transmission due to effective photon-photon interaction. We then devise a systemic method to modify the quasi-classical analysis to give much better results. 
%Lastly, we examine light propagation in an inhomogeneous lattice of side-coupled qubits and observe non-monotonicity in light transmission with increasing light intensity.
\end{abstract}

\maketitle

\section{Introduction}
\begin{figure} 
    \centering
    \includegraphics[width = \linewidth]{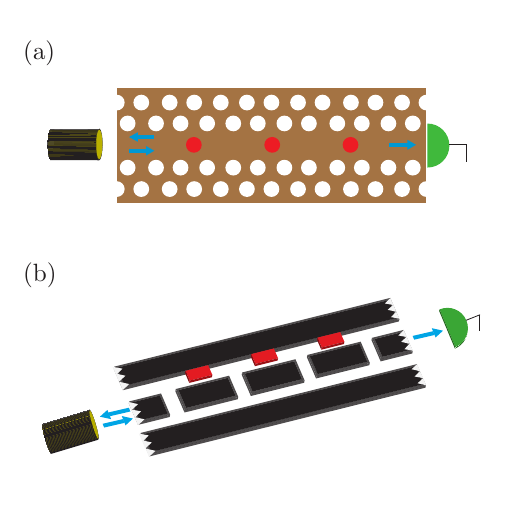}
    \caption{Cartoons showing physical realizations of the direct-coupled (a) and side-coupled (b) qubit systems. The direct-coupled qubit system can be realized using quantum dot qubits (shown as red circles) embedded in a line-defect waveguide of a photonic crystal. On the other hand, the side-coupled qubit system can be realized as an array of superconducting qubits (shown as red squares) coupled to a lattice of resonators made of transmission lines (black rectangles in the middle). We also show the sources of input photons and the detectors at the left and right of the systems respectively as well as the incident, reflected and transmitted (blue arrows) light for each system.}
    \label{NewFig1}
\end{figure}
The physics of mesoscopic collections of quantum particles has been extensively  explored in the past for investigating quantum effects such as size confinements, interference, interactions, and band-topology \cite{Akkermans2007}. The quantization of conductances, persistent currents, and Coulomb blockade are well-known examples of quantum effects in lower dimensional systems. Though these phenomena were particularly popular for mesoscopic electrical devices, recent efforts in quantum photonics have also demonstrated many interesting mesoscopic quantum effects in various cavity \cite{Walther06}, circuit \cite{Girvin09} and waveguide QED \cite{RoyRMP2017, Gu2017} set-ups. Some such recent discoveries are the photon blockade and quantum nonlinearity for propagating single photons through an elongated ensemble of laser cooled atoms in an optical-dipole trap \cite{Peyronel12, Firstenberg13}, a dissipative phase transition in a one-dimensional circuit QED lattice \cite{Fitzpatrick2017}, and the cooperative Lamb shift in a mesoscopic array of ions suspended in a Paul trap \cite{Meir14}. Motivated by these fantastic developments in experiments, we here study nonlinear quantum transport of light through one-dimensional (1D) QED lattices connected to radiation fields at the boundaries. Particularly, we investigate 1D open QED lattices either (a) in the absence of optical confinement (cavity) along the light propagation direction or (b) when the coupling to and from the cavity(ies) dominating the internal system losses in the so-called overcoupled regime \cite{RoyRMP2017, Aoki2009}. The above special features of the 1D open QED lattices separate our study from other recently explored cavity and circuit QED lattices \cite{Underwood12,Schmidt13, Raftery14, Naether15b, LeHur16,Noh16,Fitzpatrick2017,Kollar19,Orell19,Ma19}. 

%We particularly investigate effect of coherence and interaction on light propagation through a mesoscopic array of qubits. 

 %The photonic systems have their own distinctive features different from electrical devices.
 Photons being charge-neutral do not interact with each other and can have very long spatial and temporal coherence. An effective interaction between photons can be realized through their coupling to matter. Since the effective interaction is induced by the medium for photons, we also expect such interaction to affect/influence the photon transmission differently than  interaction in electron transport. Our present study employs such light-matter coupling to investigate the effects of interaction in photons' quantum transport in a 1D mesoscopic array of qubits.  These qubits can be considered to be made of alkali atoms (e.g., rubidium atoms in \cite{Peyronel12, Firstenberg13}) or superconducting qubits (e.g., transmon qubit in \cite{Fitzpatrick2017}) or ions (e.g., $Sr^{+}$ ions in \cite{Meir14}). % or dielectric ring resonators in \cite{Bandres2018}.
The features of the above qubits and their interactions with light can be very diverse.
%Nevertheless, we here consider some general models of the light-matter coupling and the interactions between qubits.
We mainly choose two different models for the 1D lattice of qubits; these are (a) a chain of qubits with direct coupling between the nearest neighbors, and  (b) a chain of connected resonators to each of which a qubit is side-coupled. While we do not have any confinement along the light-propagation for the direct-coupled qubits, we work in the overcoupled regime \cite{RoyRMP2017, Aoki2009} for the chain of connected resonators  with side-coupled qubits.
In Fig.~\ref{NewFig1}, we show representative physical implementations of the two 1D models considered here. The direct-coupled qubit system (Fig.~\ref{NewFig1}(a)) can be realized  by placing qubits such as nano-diamonds in a line-defect waveguide of a photonic crystal operating in the optical regime~\cite{Wolters10}.
The direct coupling between the qubits is generated by exchange of photons propagating through the waveguide between the qubits. Here, we integrate out these photons in the intermediate waveguide, and model the effective interaction by a constant exchange term (e.g., $J_x$) for the separation between the qubits being comparable to the wavelengths of the photons~\cite{vanLoo13}. The side-coupled qubit system is routinely realized in superconducting photonic circuits in the microwave regime~\cite{Fitzpatrick2017} (Fig.~\ref{NewFig1}(b)). A lattice of resonators can be formed by a series of capacitive gaps in the middle conductor of the 1D transmission line waveguide.

The theoretical study of correlated quantum transport of light is a nontrivial problem. Many different theoretical approaches, including analytical \cite{Shen07b,Yudson08,Yudson10,Roy10a,Zheng10,Liao10,Fan10,Koshino2012, Roy13b, Xu13,Lalumiere13,Schmidt13,Li14,Fang15,LeHur16,Hartmann16,Li16,Noh16,Roy2017} and numerical \cite{Longo10,Roy11b,Burillo14,Caneva15,Manasi2018} methods, have been explored in recent years. One major challenge in applying these theories is the scaling of the system sizes from single or multiple qubits to a mesoscopic collection of qubits.
%We here develop and implement a truncated Heisenberg-Langevin equations (THLE) approach and a quasi-classical analysis with complex interactions to scale up the system sizes.
We first study the two models using the trunacted Heisenberg-Langevin equations (THLE), which have been applied in recent years for studying correlated light transmission in a chain of two-level systems \cite{Manasi2018}. 
%We here show the benefits of truncation in the number of user variables to efficiently capture the main features of the nonlinear quantum transport and scaling up the system sizes.
Next, we compare the THLE approach to a quasi-classical analysis (QCA) \cite{Naether15} recently used in the theoretical analysis of experimental observation of dissipative phase transition in a 1D circuit QED lattice \cite{Fitzpatrick2017}. The system in \cite{Fitzpatrick2017} has relatively large qubit relaxation as well as intrinsic photon loss from every site of the medium. We are here interested in a 1D open QED system with negligible or no loss from the middle sites of it as we wish to understand the effects of interaction. It is surprising that such a QCA badly fails, in the absence of internal losses in bulk, in capturing inelastic scattering of the photons and the related effective interaction between photons generated in our models. We specifically notice that contrary to the results from the THLE, there is always some frequency that has perfect transmission within the QCA.
%It appears that the source of such discrepancy is due to failure of QCA in capturing inelastic scattering of the photons and the related effective interaction between photons generated in our models.
The main contribution of our work is a systemic modification of the interaction parameter to the complex plane such that we revive the QCA even in the absence of bulk dissipation. We develop a simple scheme to find the complex interaction parameter for the modified QCA, which only uses single qubit analysis. Quite remarkably, the modified QCA works quite well when it is compared to the more accurate THLE even for larger systems. Unlike the QCA, the modified QCA is able to cause dissipation in the medium through the complex interaction parameter. Which gives accurate results even at higher intensities at a much less computational cost than the THLE. %The advantage of the modified QCA is that it is simpler to implement numerically and has very low computational cost for much longer lattices than the THLE. 

\begin{figure} 
    \centering
    \includegraphics[width = \linewidth]{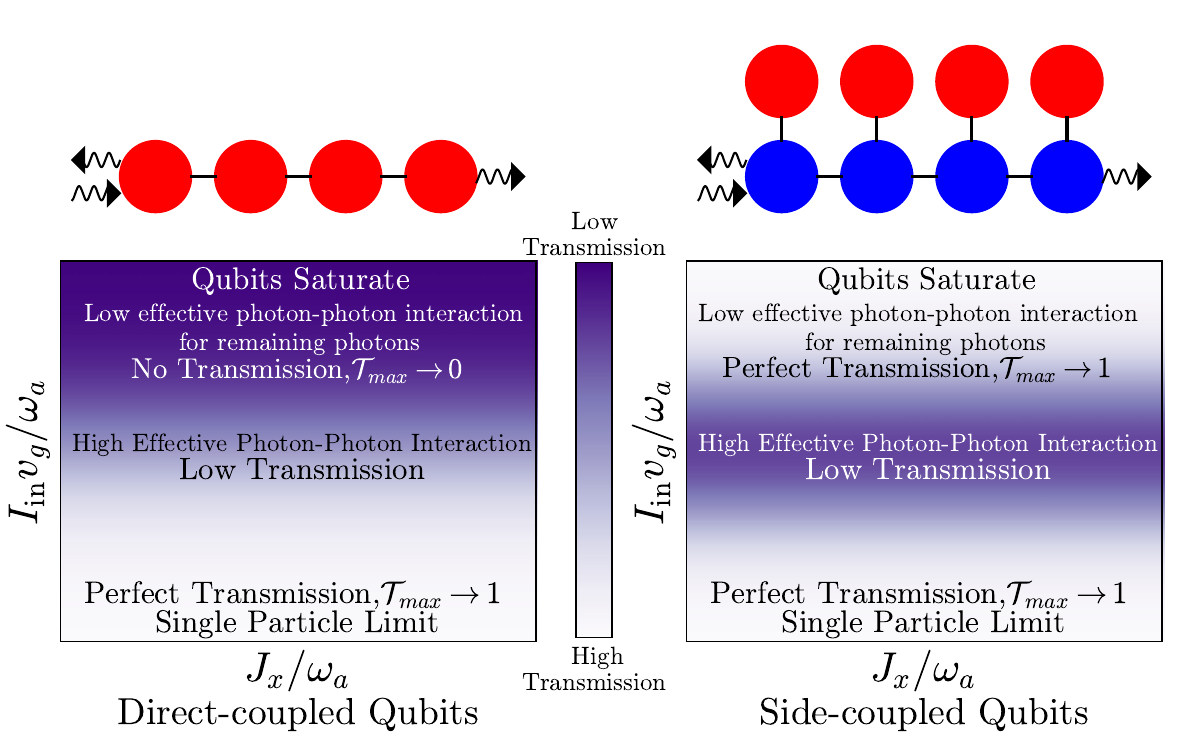}
    \caption{Schematic diagrams showing the arrangement of the direct and side-coupled qubits mediums through which the photons are incoming and getting reflected on the left and being transmitted to the right. The red circles represent the qubits while the blue ones represent the resonators. For each medium we show, in the parameter space of input intensity~($I_{\rm in}$) and inter-site coupling~($J_x$), the regions of high transmission with the color gradient. With increasing intensity we observe an initial increase in effective photon-photon interaction followed by the saturation of qubits at very high intensity. The saturation of qubits has opposite effect on transmission in the two mediums, causing a complete reduction of transmission in the direct-coupled system while a revival of full resonant transmission through the resonators for the side-coupled one.}
    \label{fig:FigI1}
\end{figure}

%We then apply the THLE and modified QCA to study the effects of interaction in light propagation through the two 1D open QED lattice models. 
At low intensities of the incident light, we find perfect resonant transmission in both the models, which can be investigated by any of THLE, QCA, and modified QCA. The interaction dominates light propagation for increasing intensities in an array of direct-coupled qubits. The transmission probability in this model falls with an increasing power due to the photon-photon blockade, which is generated by the nonlinearity (interaction) at the qubits. This regime can be adequately analyzed only by either THLE or modified QCA. For the side-coupled qubits, the effect of interaction on light propagation is non-monotonic with increasing intensity of input light. The light transmission probability at resonant condition first decays with growing power, and then it again enhances with further increase in intensity when photons saturate the qubits. Our work provides an insightful understanding of the role of excitation blockade on the transmission by proposing a simple calculation method to capture their effects. The rest of the paper is organized in the following sections and appendices. The models, analysis, primary results, and outlook are given in the next four sections. 
%We also include technical details of our research in the appendix for the interested readers. 
We include the details of calculations, evidences of inelastic scattering, results for inhomogeneous side-coupled system, etc in the appendices.
\section{Models}

%In this work we focus on the inelastic scattering of the photons phenomena that is caused due to interactions. We consider the 1D system either comprising of qubits with  direct coupling between the nearest neighbors or a resonator chain to which the qubits are side-coupled.

To study the transmission of light through the system, we shine light from one end (left) using a laser or a microwave generator depending on the experimental system and observe them at the other end of the system with a suitable detector. In the physical realizations of these systems, the waveguides made of line-defect photonic crystals or transmission lines are much longer than the typical wavelength of photons employed in that experiment. Thus, the waveguide can be treated as infinitely long. Nevertheless, the length of the systems in our effective theoretical modeling is finite and is the same as the number of qubits or resonators. In our studied Hamiltonians, the sources and detectors are modeled as the left and right baths respectively. %In the physical systems there may also be small waveguides connecting the qubits or the resonators. %The intermediate waveguides result in non-trivial interaction between the qubits or resonators by exchange of real or virtual photons when the distance between the qubits is comparable to the wavelengths of the photons~\cite{vanLoo13}.

The entire system including the baths is described by the Hamiltonian

\begin{align}
    H = H_M + H_{LB} + H_{RB} + H_{LM} + H_{RM},
\end{align}  
where $H_M$ represents the Hamiltonian of the medium which, in this paper, can be the direct-coupled qubits ($H_{direct}$) or resonator chain with side-coupled qubits ($H_{side}$). $H_{LB}$ ($H_{RB}$) is the left (right) bath and $H_{LM}$ ($H_{RM}$) is the coupling between the chain and the left (right) bath. The bath and the bath-medium coupling Hamiltonians within the Markov and rotating-wave approximation are \cite{Manasi2018}
\begin{align}
    H_{LB(RB)} &= \int_{-\infty}^{+\infty} dk\:\hbar \omega_{k}a_{L k(R k)}^{\dagger}a_{L k(R k)}, \\
    %H_{RB} &= \int_{-\infty}^{+\infty} dk\:\hbar \omega_{k}a_{R k}^{\dagger}a_{R k},\\ 
    H_{LM} &= \int_{-\infty}^{+\infty} dk\: \hbar g_{L} (a_{L k}^{\dagger}\chi_{1} + \chi_{1}^{\dagger}a_{L k}), \\
H_{RM} &= \int_{-\infty}^{+\infty} dk\: \hbar g_{R}(a_{R k}^{\dagger}\chi_{N} + \chi_{N}^{\dagger}a_{R k}). 
\end{align}
 The operators $a_{L k}^{\dagger}$ and $a_{R k}^{\dagger}$ are the creation operators with wave number $k$ in the left and right side baths respectively. We set the ground state energy of the bath to be zero and have $[ a_{L k'},a_{L k}^{\dagger}] = [a_{R k'},a_{R k}^{\dagger}] =  \delta(k-k^{'})$. The bosonic operators $\chi_{1}$ and $\chi_{N}$ correspond to the left and right most qubits/resonators of the medium, which are coupled to the baths with strengths $g_L$ and $g_R$ respectively. We use $\chi_{j}=b_j~(f_j)$ for a direct-coupled (side-coupled) array of qubits (see below). Both the baths contain left- and right-moving photons; hence $k$ can take both positive and negative values representing photons moving towards right and left respectively. However, because of our particular initial condition only right-moving photons for the right baths are relevant. Here, we assume a linear energy-momentum dispersion relation for the two baths, i.e., $\omega_k = v_g |k|$ with $v_g$ being the group velocity of photons. The $k$-independent couplings $g_L$ and $g_R$ within the Markov approximation ensure that we can turn a set of integro-differential Heisenberg-Langevin equations for qubit variables into a set of differential equations  (see Sec.~\ref{ModelAnalysis}).    

 The baths are kept at a temperature of 0K to ensure a coherent light source. At time $t_0$ the medium-bath coupling is switched on and incoming light interacts with the ends of the chain. For a weak bath-medium coupling, we write the $H_{LM}$ and $H_{RM}$ within the rotating wave approximation where we drop the counter rotating terms. Assuming that a coherent light is incident from the left and the initial state (at $t = t_0$) for the coherent light source is $|E_p, \omega_p \rangle$,
\begin{align}
    a_{L k}(t_0)|E_p,\omega_p\rangle &= E_p\delta(v_gk-\omega_p)|E_p,\omega_p\rangle, \label{initialcond}\\
    a_{R k}(t_0)|E_p,\omega_p\rangle &= 0, \label{initialcond2}
\end{align}
where $E_p$ is the amplitude of the incoming light while $\omega_p$ is its frequency. The initial condition converts the noises from the bath into the Rabi frequency of the system, given by $\Omega_L=g_LE_p/v_g$, and medium-bath coupling rates from the left and right baths given by, $\Gamma_L=\pi g_L^2/v_g$ and $\Gamma_R=\pi g_R^2/v_g$ respectively. The noise terms can now be replaced in the quantun Langevin equations for the qubits/resonators with these constant terms. The Rabi frequency appears in the system of equations either as a constant source term or as prefactors to certain operators, while the bath-medium coupling rates only appear as the latter. The incoming photon frequency~($\omega_p$) appears as time-dependent phase factors in the quantum Langevin equations which are absorbed by suitable redefinition of the operators.

\subsection{Direct-coupled array of qubits}
A simple Bose-Hubbard type model system that shows correlated light propagation is a 1D array of qubits with direct coupling between nearest neighbors \cite{Schmidt13,Naether15b,Ma19}:
\begin{align}
     H_M=H_{direct} = &\sum_{j=1}^{N}\hbar\omega_{q_j}b_j^{\dagger}b_j + \hbar Ub_j^{\dagger}b_j(b_j^{\dagger}b_j - 1)
     \nonumber \\ \quad &+ \sum_{j=1}^{N-1}2\hbar J_x(b_j^{\dagger}b_{j+1} + b_{j+1}^{\dagger}b_j).  
\end{align}
Here, $b^{\dagger}_j$ and $b_j$ are the bosonic creation and annihilation operators at site $j$ with $\omega_{q_j}$ being the qubit frequency and $[b_{i}, b_j^{\dagger}]=\delta_{i,j}$. The direct-coupled medium is coupled to the two baths through $\chi_{1}=b_{1}$ and $\chi_{N}=b_{N}$. $2J_x$ is the coupling strength between neighboring qubits. $U$ determines the strength of on-site interaction, which is repulsive for positive $U$. In the limit $U \to \infty$, the qubit maps to a two-level system. The on-site interaction at the qubits induces optical nonlinearity in the model by preventing multiple excitations by photons, and is responsible for the inelastic light scattering studied in this work.

\subsection{Side-coupled array of qubits} \label{SCQ_Model}
The physical system investigated in \cite{Fitzpatrick2017} corresponds to a chain of connected resonators to each of which a qubit is side-coupled. In contrast to \cite{Fitzpatrick2017}, we consider here no photon loss from the resonators in the one-dimensional open QED medium. The Hamiltonian of the model is
\begin{align}
    H_M = H_{side} = &\sum_{j=1}^N (H_j^q + H_j^r + H_j^{rq}) + \sum_{j=1}^{N-1} H^{hop}_{j},  
\end{align}  
where,
\begin{align*}
    H_j^q &= \hbar \omega_{q_{j}} b_j^{\dagger}b_j + \hbar Ub_j^{\dagger}b_j(b_j^{\dagger}b_j-1), \\ 
    H_j^r &= \hbar\omega_{r_{j}} f_j^{\dagger} f_j, \qquad
    H_j^{rq} = \hbar g_j(b_j^{\dagger}f_j + f_j^{\dagger}b_j), \\ 
    H^{hop}_{j} &= 2\hbar J_x(f_j^{\dagger}f_{j+1} + f_{j+1}^{\dagger}f_j).
\end{align*}
Similar to the directly coupled qubit chain, $b_j^{\dagger}$ and $b_j$ are the creation and annihilation operators for the qubits with qubit frequency $\omega_{q_j}$ and on-site interaction $U$. The resonator chain is a 1D array of bosonic resonators with creation and annihilation operators $f_j^{\dagger}$ and $f_j$, respectively, and resonator frequency $\omega_{r_j}$. Each resonator in the chain is coupled to its nearest neighbors, with coupling strength $2J_x$, and side-coupled with the qubit, with resonator-qubit coupling strength $g_j$. The side-coupled medium is coupled to the two baths through $\chi_{1}=f_{1}$ and $\chi_{N}=f_{N}$.

\section{Analysis of the models} \label{ModelAnalysis}
\subsection{Integrating out the baths}
Following ~\textcite{Manasi2018}, we first integrate out the baths. The Heisenberg equations for the photon operators of the bath, $a_{L k}$ and $a_{R k}$ are inhomogeneous linear differential equations. The initial condition (at $t=t_0$) will set the direction of the incoming light in the system. By solving the Heisenberg equations for the baths, we obtain the time evolution of the bath operators~\cite{Manasi2018} given by,
\begin{eqnarray} 
    a_{L k}(t)&=&\mathcal{G}_k(t-t_0)a_{L k}(t_0) - ig_L\int_{t_0}^tdt'\mathcal{G}_k(t-t')\chi_1(t'),\nonumber \\\label{batheqnL}\\
    a_{R k}(t)&=& \mathcal{G}_k(t-t_0)a_{R k}(t_0) - ig_R\int_{t_0}^tdt'\mathcal{G}_k(t-t')\chi_N(t'),\nonumber \\ \label{batheqnR}
\end{eqnarray}
where, $\mathcal{G}_k(\tau)=\theta(\tau)e^{-iv_gk\tau}$ is the retarded Green's function of the baths. The bath solutions, Eqs.~\ref{batheqnL}-\ref{batheqnR}, are substituted in the remaining equations of motion for the qubits and give rise to the noise operators due to the baths which are given as, $ \eta_L(t) = \int^{\infty}_{-\infty} dk\: \mathcal{G}_k(t-t_0)g_La_{L k}(t_0)$ and  $\eta_R(t) = \int^{\infty}_{-\infty} dk\: \mathcal{G}_k(t-t_0)g_Ra_{R k}(t_0)$. These noises coming from the baths can be converted to constant numbers by considering an appropriate initial condition.
\subsection{Truncated Heisenberg-Langevin equations} \label{THLEAnalysis}
As the qubits (resonators) are bosonic, each of them can have any number of excitations. These excited states appear as powers of $b_j$ and $b_j^{\dagger}$ ($f_j$ and $f_j^{\dagger}$) in the operators. To describe the THLE, we need to consider every independent boson operator for each qubit/resonator for each site; which are $\{I, b_j,b^{\dagger}_j, b^{\dagger}_jb_j, b_j^2, b_j^{\dagger 2}, b_j^{\dagger}b_j^2, b_j^{\dagger 2}b_j, b_j^{\dagger 2}b_j^2,... \}$. However, as there are infinite number of possible operators, we must truncate the set of boson operators. If we restrict our system to have only one excitation, then only the first four of these operators would enter the equations. In this case, for example, to obtain the set of equations for two qubits we consider every operator in the set $\{I, b_1, b_1^{\dagger}, b_1^{\dagger}b_1\}\otimes\{I, b_2, b_2^{\dagger}, b_2^{\dagger}b_2\}$, which gives 16 independent operators. The operator $I\otimes I$ is ignored as it has no dynamics and we are left with 15 operators. The number of operators increases exponentially as number of qubits $(N)$ increases and are given by $\{I, b_j, b_j^{\dagger}, b_j^{\dagger}b_j\}^{\otimes N}$.

 Including more excited states in our operator set will increase the accuracy of the results, however, for relatively high on-site interaction $(U)$, appropriate accuracy can be achieved by considering a limited number of excited states. As more excited states per site $(m)$ are included, the number of operators increases with a polynomial growth where the exponent is $2N$, we denote $m$ as maximum number of bosons per site. For example, if $m=2$, the number of operators for a chain of two qubits ($N=2$) increases to 80 as the operator set becomes $\{I, b_j,b_j^{\dagger},  b_j^{\dagger}b_j, b_j^2, b_j^{\dagger 2}, b_j^2b_j^{\dagger}, b_jb_j^{\dagger 2}, b_j^2b_j^{\dagger 2} \}^{\otimes 2}$. Similarly, for a chain with $N$ qubits with $m$ allowed excited states per site, we shall get $(m+1)^{2N}-1$ operators.
 
We illustrate the THLE analysis by writing the equations of motion for the simplest system considered here, namely a single qubit. The expectation values of qubit operators after integrating the baths from the equations along with its phase factor are denoted as,
\begin{equation} \label{Sjk}
    \mathcal{S}_{kl} = \langle b_1^{\dagger k}b^l_1  \rangle e^{(-k+l)i\omega_p(t-t_0)},
\end{equation}
where the phase factor has been chosen such that the RHS of the equations of motion have no explicit time dependence. A few examples can be, $\mathcal{S}_{01} = \langle b_1 \rangle e^{i\omega_p(t-t_0)}$, $ \mathcal{S}_{22} = \langle b_1^{\dagger 2}b^2_1 \rangle$. 
Applying this convention to our equations, and setting $m=2$, we get the following set of eight coupled differential equations:

\begin{eqnarray}
    \mathcal{\dot S}_{01}&=&-(i\delta\omega_{q_1} + \Gamma_L + \Gamma_R)\mathcal{S}_{01} - 2iU\mathcal{S}_{12} - i\Omega_L, \label{S01}\\
    \nonumber\\
    \mathcal{\dot S}_{02}&=&-(\delta\omega_{q_1} + 2\Gamma_L + 2\Gamma_R)\mathcal{S}_{02} - 2iU(\mathcal{S}_{02} + 2\mathcal{S}_{13}) \nonumber \\& &- 2i\Omega_L\mathcal{S}_{01}, \label{S02}\\
    \nonumber\\
    \mathcal{\dot S}_{11}&=& -2(\Gamma_L + \Gamma_R)\mathcal{S}_{11} - i\Omega_L(\mathcal{S}_{10} - \mathcal{S}_{01}),\label{S11}\\
    \nonumber \\
    \mathcal{\dot S}_{12} &=&-(i\delta\omega_{q_1} + 3\Gamma_L + 3\Gamma_R)\mathcal{S}_{12} - 2iU(\mathcal{S}_{12} + \mathcal{S}_{23})  \nonumber \\& &- i\Omega_L(2\mathcal{S}_{11} - \mathcal{S}_{02}), \label{S12}\\
    \nonumber \\
    \mathcal{\dot S}_{22} &=& 2( 2\Gamma_L + 2\Gamma_R)\mathcal{S}_{22} - i\Omega_L(2\mathcal{S}_{21}-2\mathcal{S}_{12}). \label{S22}
\end{eqnarray}
here, $\delta\omega_{q_1} = \omega_{q_1} - \omega_p$ and the remaining equations are the hermitian conjugates of the equations \ref{S01}, \ref{S02} and \ref{S12}. In the truncation scheme with $m=2$, we ignore all other equations with higher values of $k$, $l$ in Eq.~\ref{Sjk}. Note that we still get the operators we have ignored due to the truncation on the RHS of the above equation for example in Eqs.~\ref{S02}-\ref{S12} and their hermitian conjugates. We drop these terms in the further analysis, and explicitly check the accuracy of the truncation by considering various values of $m$. As mentioned before, increasing $m$ increases accuracy of the results but also expands the set of equations.

The equations \ref{S01}-\ref{S22}, and their hermitian conjugates, can be re-written in a matrix form with the vectors $\boldsymbol{\mathcal{S}}(t)$ = $\{ \mathcal{S}_{01}, \mathcal{S}_{02}, \mathcal{S}_{10}, \mathcal{S}_{11}, \mathcal{S}_{12}, \mathcal{S}_{20}, \mathcal{S}_{21}, \mathcal{S}_{22} \}^T$ and $\boldsymbol{\Omega} = \{-i\Omega_L,0,i\Omega_L,0,0,0,0,0 \}^T$. The linear system of equations can be represented very compactly as

\begin{align} \label{mateq}
    \boldsymbol{\mathcal{\dot S}(t)} = \boldsymbol{\mathcal{Z}}\boldsymbol{\mathcal{S}}(t) + \boldsymbol{\Omega}.
\end{align}

We analyse both models using the Heisenberg-Langevin equations which is an exact calculation method for an open quantum system within the Markovian approximation \cite{scully_zubairy_1997,RoyRMP2017,Manasi2018}. 
%The method is equivalent to quantum master equation. 
%As we are modelling 
Since each component is modeled as harmonic or anharmonic oscillators the local Hilbert space is infinite dimensional. As a result we necessarily have to perform a truncation in any numerical calculation.
This truncation is a further approximation whose correctness is checked explicitly by considering various degree of truncation. Assuming the qubits in the ground state before irradiation with photon, we solve the coupled set of equations to get the steady-state results, which occurs when $t \gg t_0$ after some transient time. We obtain the long-time $(t\to \infty)$ steady-state solution by solving $\frac{d\boldsymbol{\mathcal{S}}(t)}{dt}=0$, which gives the time independent solution of the equations as $\boldsymbol{\mathcal{S}}(t \to \infty) = -\boldsymbol{\mathcal{Z}}^{-1}\boldsymbol{\Omega}$. 
%Master equation solutions are the most known method to deal with equation of motion for evolution of quantum systems using schrodinger equation in schrodinger picture. If context demands to use Heisenberg picture approach then called quantum Heisenberg-Langevin approach and after truncating the equation we get Truncated Heisenberg-Langevin equations. As we increase the system sizes further in the THLE analysis, the number of THLEs grows exponentially which tends to reach the computational limit even after large truncation, (see appendix).
\subsection{Integrating out the resonators} \label{ResonatorIntegration}

To model the side-coupled system of qubits, we first solve the equations for the resonators coupled to the baths and qubits, just as it was done in the case of baths. For a single qubit-resonator pair, we denote the expectation values of resonator operators after integrating the baths similar to those of the qubits in Eq.~\ref{Sjk} by:
\begin{equation} \label{Fjk} 
    F_{kl}(t)= \langle f_1^{\dagger k}f^l_1  \rangle e^{(-k+l)i\omega_p(t-t_0)}.
\end{equation}
With this notation, the equation for the resonator variable becomes the following  equation:
\begin{equation} \label{F01}
    \dot F_{01}(t)=-(i\delta\omega_{r_1}+\Gamma)F_{01}(t) - ig_{1}\mathcal{S}_{01}(t)-i\Omega_L.
\end{equation}
Here, $\delta\omega_{r_1}=\omega_{r_1} - \omega_p$, $\Gamma=\Gamma_L+\Gamma_R$ and $\mathcal{S}_{01}(t)$ is the qubit operator as mentioned in Sec.~\ref{THLEAnalysis}.  Since the resonator has no non-linearity (it is modeled as a harmonic oscillator), we note that Eq.~\ref{F01} is an inhomogeneous linear differential equation which is not connected to any other variables of the resonator apart from $F_{01}$. We now integrate Eq.~\ref{F01} to obtain a formal solution of the equation as:
\begin{align} \label{solvedF01}
    F_{01}(t)=& F_{01}(t_0)\mathcal{G}_A(t-t_0) - i\Omega_L \int_{t_0}^{t}\mathcal{G}_A(t-t^{\prime})dt^{\prime} \nonumber \\
    &-ig_{1}\int_{t_0}^{t}\mathcal{G}_A (t-t^{\prime}) \mathcal{S}_{01}(t^{\prime})dt^{\prime},
\end{align}
where, $\mathcal{G}_A(\tau)=e^{-A\tau} \theta (\tau)$ and $A=i\delta\omega_{r_1}+\Gamma$. This is quite analogous to integrating out the baths, except that the last integral in Eq.~\ref{solvedF01} retains a memory of the  qubits' time-evolution; thus, the Eq.~\ref{solvedF01} is non-Markovian in nature. For simplicity, we now make the Markovian approximation, and we replace $\mathcal{S}_{01}(t^{\prime})$ in Eq.~\ref{solvedF01} by $\mathcal{S}_{01}(t)$. This simplifies the equation by neglecting any retardation effect. We can now write at long-time $(t\to \infty)$ steady-state: %To obtain the steady-state solutions of Eq.~\ref{solvedF01}, we solve for the steady-state when $\mathcal{S}_{01}$ becomes constant in time. With this assumption, the $\mathcal{S}_{01}$ in the integrand can be taken to be a constant (its steady-state value). This is because the Green's function ($\mathcal{G}_A(\tau)$) decays exponentially and the transient values of $\mathcal{S}_{01}$ make exponentially small negligible contribution to the integral. Thus, to obtain the steady-state solution, we may take $\mathcal{S}_{01}$ out of the integral and perform the rest of the integration. The system of equations obtained in this manner is valid at long times and can be used to calculate the steady-state. With this assumption, at long times the Eq.~\ref{solvedF01} becomes
\begin{equation} \label{longTF01}
    F_{01}(t \rightarrow \infty) = -i\Omega_L / A - ig_{1}\mathcal{S}_{01}(t \rightarrow \infty)/A.
\end{equation}

The Eq.~\ref{longTF01} is then substituted in the differential equations obtained after taking expectation of the equations of the qubits, which for $m$=2 at long time $(t\to \infty)$ gives

\begin{eqnarray}
    \mathcal{\dot S}_{01}( \infty) &=& -\left[i\delta\omega_{q_1} + g_{1}^2/A \right]\mathcal{S}_{01}-2iU\mathcal{S}_{12} - g_{1}\Omega_L/A, \nonumber \\\label{QubitS01}\\
    \mathcal{\dot S}_{02}(\infty) &=& -2\left[i\delta\omega_{q_1} + g_{1}^2/A \right]\mathcal{S}_{02} - 2iU(\mathcal{S}_{02} + 2\mathcal{S}_{13}) \nonumber \\& &- 2g_{1}\Omega_L\mathcal{S}_{01}/A, \label{QubitS02}\\
    \mathcal{\dot S}_{11}(\infty) &=& -\left[g_{1}^2/A+g_{1}^2/A^* \right]\mathcal{S}_{11} - g_{1}\Omega_L\mathcal{S}_{10}/A \nonumber \\& &- g_{1}\Omega_L\mathcal{S}_{01}/A^*, \label{QubitS11}\\
    \mathcal{\dot S}_{12}(\infty) &=&-\left[ i\delta\omega_{q_1} + 2g_{1}^2/A + g_{1}^2/A^* \right]\mathcal{S}_{12}  \nonumber \\& &- 2iU(\mathcal{S}_{12}+2\mathcal{S}_{23}) 
     -g_{1}\Omega_L 2\mathcal{S}_{11}/A \nonumber \\& &-g_{1}\Omega_L \mathcal{S}_{02}/A^*, \label{QubitS12}\\
    \mathcal{\dot S}_{22}(\infty) &=& -2\left[g_{1}^2/A+g_{1}^2/A^* \right]\mathcal{S}_{22} - 2g_{1}\Omega_L \mathcal{S}_{21}/A \nonumber \\& &- 2g_{1}\Omega_L  \mathcal{S}_{12}/A^*, \label{QubitS22}
\end{eqnarray}
and the rest of the equations are hermitian conjugates of \ref{QubitS01}, \ref{QubitS02}, \ref{QubitS12}. The Eqs.~\ref{QubitS01}-\ref{QubitS22} and its hermitian conjugates can be re-written in the matrix form and solved to obtain the steady-state solution in the same way as mentioned in Sec.~\ref{THLEAnalysis}.

For longer system lengths, similar to the system of single qubit-resonator pair, we obtain $N$ linear inhomogeneous equations for the resonators in an array of $N$ resonator-qubit pairs. Also similar to the single site case, the above mentioned approximations can be applied to integrate the resonators' equations and to obtain the long-time steady-state relations between $F$'s and $\mathcal{S}$'s. After integration, the steady-state relations between $F$'s and $\mathcal{S}$'s are in turn substituted in the differential equations obtained after taking expectation of the equations of the qubits. As the resonators can be integrated out, the numerical complexity of the problem is determined only by the qubits.

\subsection{The quasi-classical analysis}

The models studied here can also be investigated using a quasi-classical system of equations, which are obtained by the classical limit approximation  where the operators are replaced by their averages $( \langle b_j \rangle=\beta_j)$ for e.g. $\langle b_j^{\dagger}b_j^2 \rangle = |\beta_j|^2\beta_j$ \cite{Naether15}. This approximation limits the set of equations in their first moment. For the direct-coupled qubits system, we obtain
\begin{align} \label{QC_direct}
    \dot{\beta_j} = &-(i\delta\omega_{q_j} + \Gamma_j)\beta_j - 2iU|\beta_j|^2\beta_j \nonumber \\
    &- 2iJ_x(\beta_{j-1} + \beta_{j+1}) -i\Omega_L\delta_{1,j},
\end{align}
where, $\Gamma_j=0$ for $2 \leq j \leq N-1$, $\Gamma_1=\Gamma_L$ and $\Gamma_N=\Gamma_R$. If $N=1$, $\Gamma = \Gamma_L+\Gamma_R$. We use open boundary condition for the medium, i.e., $\beta_0=\beta_{N+1}=0$.

The quasi-classical equations can be derived for the side-coupled qubits in the same manner. For the resonator chain, let $ \langle f_j \rangle=\alpha_j$ and hence for $N$ resonator-qubit pairs in the side-coupled model we get $2N$ quasi-classical equations which are
\begin{align} 
    \dot{\alpha_j} = &-(i\delta\omega_{r_j}+\Gamma_j)\alpha_j - 2iJ_x(\alpha_{j-1} + \alpha_{j+1}),\label{QC_sidecoupledA} \nonumber \\
    &-ig_j\beta_j -i\Omega_L\delta_{1,j} \\
    \dot{\beta_j} = &-i\delta\omega_{q_j}\beta_j - 2iU|\beta_j|^2\beta_j - ig_j\alpha_j,\label{QC_sidecoupledB}
\end{align}
where, $\Gamma_j=0$ for $2 \leq j \leq N-1$, $\Gamma_1=\Gamma_L$ and $\Gamma_N=\Gamma_R$. If $N=1$, $\Gamma = \Gamma_L+\Gamma_R$. Again we use the open boundary condition, i.e. $\alpha_0=\alpha_{N+1}=0$.

 These equations are solved using standard numerical differential equation solvers to obtain the long-time steady-state solutions. At very low intensities when the interaction and non-linearity in the system are not significant, the QCA has a perfect agreement with the THLE results. However, with increasing intensity, the QCA fails to capture any effects of inelastic scattering of the photons as predicted by the THLE analysis. The peaks in transmission profile calculated using QCA starts shifting towards higher frequencies as the intensity is increased. This drastic failure of QCA is possibly a reflection of the fact that the approximation is missing the inelastic scattering of the photons due to interaction. If we consider the Eq.~\ref{QC_direct} carefully we realize that the interaction term is contributing as extra occupation dependent contribution to the oscillation frequency instead of a term that can cause a decrease in oscillation amplitude. The blockade phenomena should have contributed the latter. In particular, for the QCA, there is always some frequency that has perfect transmission. In the case where input intensity is very high, the shift can be so large that the peak may go outside the range of input frequencies considered and it can be mistakenly interpreted as very low transmission caused by interaction. 
 
The non-linearity in the QCA due to on-site interaction term can cause more than one steady-states, e.g., a multi-stability in the system which is absent in the present THLE analysis with one-time variables. It is also possible to find stable oscillating solutions for these quasi-classical equations. These non-linear features of the quasi-classical equations appear at high intensities, while the QCA begins to fail even at much lower intensity. We also carried out a self-consistent mean-field analysis of the system (see Appendix~\ref{SCMF}) but it also fails to capture the inelastic scattering of the photons at higher intensities. Interestingly, both the approximate analyses, namely QCA and self-consistent mean-field, give the same incorrect results for a wide range of parameters.

\subsection{Modified quasi-classical equations} \label{secUeff}
In contrast to THLE analysis, the quasi-classical equations can be solved very easily as there are only $N$ equations involved. The quasi-classical equations increase linearly with system size $N$ and hence enables us to analyze number of system with hundreds of qubits. A careful analysis of the quasi-classical equations reveals that the above mentioned shift in the transmission profile is due to the on-site interaction term $(U)$. We observe that we need to incorporate inelastic scattering of the photons and related reduction in the transmission within the QCA. To highlight the fact that this inelastic scattering of the photons and the related reduction in photon transmission are caused due to interaction, we modify the interaction parameter $U$ to become a complex number ($U_{\rm eff}$). We choose its value to match the THLE analysis results for the single qubit case. 

To calculate $U_{\rm eff}$ for the direct-coupled qubits, we consider a truncated system of five operators $S^T = \left[\begin{matrix}\mathcal{S}_{01} & \overline{\mathcal{S}_{01}} & \mathcal{S}_{11} & \mathcal{S}_{12} & \overline{\mathcal{S}_{12}}\end{matrix}\right]$ and calculate their THLE which are given by Eqs. \ref{S01}, \ref{S11}, \ref{S12} and the conjugates of Eqs. \ref{S01}, \ref{S12}.

By solving Eqs.~\ref{S11} and \ref{S12}, we obtain the steady-state values $\mathcal{S}^{\infty}_{11}$ and $\mathcal{S}^{\infty}_{01}$ in terms of $\mathcal{S}^{\infty}_{01}$
\begin{equation}
\mathcal{S}^{\infty}_{11} = \frac{i \Omega_L \left(\mathcal{S}^{\infty}_{01} - \overline{\mathcal{S}^{\infty}_{01}}\right)}{2 \Gamma},
\end{equation}
\begin{equation}
\mathcal{S}^{\infty}_{12} = \frac{\Omega_L^{2} \left(\mathcal{S}^{\infty}_{01} - \overline{\mathcal{S}^{\infty}_{01}}\right)}{\Gamma \left(i (\delta\omega_{q_1}+2U) + 3 \Gamma\right)},
\end{equation}
where, $\Gamma = \Gamma_L + \Gamma_R$. The steady-state solutions are substituted in Eq.~\ref{S01} at steady-state which gives
\begin{equation} \label{appS01ss}
\mathcal{S}^{\infty}_{01} \left(- \Gamma - i \delta\omega_{q_1}\right) - \frac{2 i U \Omega_L^{2} \left(\mathcal{S}^{\infty}_{01} - \overline{\mathcal{S}^{\infty}_{01}}\right)}{\Gamma \left(i (\delta\omega_{q_1}+2U) + 3 \Gamma \right)} - i \Omega_L = 0,
\end{equation}

The Eq.~\ref{appS01ss} can be re-written in real and imaginary parts of $\mathcal{S}^{\infty}_{01}$ by substituting $\mathcal{S}^{\infty}_{01} = S_x + iS_y$ which gives two equations for the two real variables. We solve these two equations to obtain steady-state values of $S_x$ and $S_y$ as Eqs.~\ref{S_x} and \ref{S_y}.
\begin{widetext}
\begin{equation}
S_x = - \frac{\Omega_L \left((\delta\omega_{q_1}+2U)^{2} \delta\omega_{q_1} + 12 U \Omega_L^{2} + 9 \Gamma^{2} \delta\omega_{q_1}\right)}{(\delta\omega_{q_1}+2U)^{2} \left(\Gamma^{2} + \delta\omega_{q_1}^{2}\right) + 4 (\delta\omega_{q_1}+2U) U \Omega_L^{2} + 12 U \Omega_L^{2} \delta\omega_{q_1} + 9 \Gamma^{4} + 9 \Gamma^{2} \delta\omega_{q_1}^{2}} \label{S_x}
\end{equation}
\begin{equation}
S_y = - \frac{\Gamma \Omega_L \left((\delta\omega_{q_1}+2U)^{2} + 9 \Gamma^{2}\right)}{(\delta\omega_{q_1}+2U)^{2} \left(\Gamma^{2} + \delta\omega_{q_1}^{2}\right) + 4 (\delta\omega_{q_1}+2U) U \Omega_L^{2} + 12 U \Omega_L^{2} \delta\omega_{q_1} + 9 \Gamma^{4} + 9 \Gamma^{2} \delta\omega_{q_1}^{2}}.  \label{S_y}
\end{equation}
\end{widetext}

For comparison with the QCA at steady-state, the quasi-classical equation for the medium of single qubit is given as
\begin{equation} \label{betass1}
- 2 i U |\beta_{1}^{ss}|^{2} \beta_{1}^{ss} - i \Omega_L + \beta_{1}^{ss} \left(- \Gamma - i \delta\omega_{q_1}\right) = 0.
\end{equation}
The transmission probability for the THLE and QCA for $N=1$ at steady-state is given by
\begin{align}
    \mathcal{T}_{\rm THLE} &= \frac{2\Gamma_R}{v_g I_{\rm in}}\mathcal{S}_{11}^{\infty}, \\
    \mathcal{T}_{\rm QCA} &= \frac{2\Gamma_R}{v_g I_{\rm in}}|\beta_{1}^{ss}|^2.
\end{align}
In order to match these transmission probabilities for the THLE steady-state obtained above with the QCA steady-state, we obtain the magnitude of $\beta_{1}^{ss}$ by demanding that the transmission obtained in the two calculations are the same, which gives
\begin{equation}
|\beta_{1}^{ss}|^2 = \mathcal{S}^{\infty}_{11} = - \frac{S_{y} \Omega_L}{\Gamma}.
\end{equation}
 Comparing transmission, however, does not give us the phase of $\beta_{1}^{ss}$ but we choose it to be the same as that of $\mathcal{S}^{\infty}_{01}$ which yields
\begin{equation}  \label{betassri}
\beta^{ss}_{1,r} = \frac{|\beta^{ss}_{1}| S_x}{\sqrt{S_x^2+S_y^2}} \qquad \beta^{ss}_{1,i} = \frac{|\beta^{ss}_{1}| S_y}{\sqrt{S_x^2+S_y^2}},
\end{equation}
where $\beta^{ss}_{1,r}$ and $\beta^{ss}_{1,i}$ are the real and imaginary parts of $\beta^{ss}_{1}$ and $|\beta^{ss}_1|=\sqrt{(\beta^{ss}_{1,r})^2 + (\beta^{ss}_{1,i})^2}$.
 
Finally, we re-define the $U$ as $U_{\rm eff}$ and re-arrange Eq.~\ref{betass1} to obtain
\begin{equation} 
U_{\rm eff} = \frac{- \Omega_L + \beta_{1}^{ss} \left(i \Gamma - \delta\omega_{q_1}\right)}{2 |\beta_{1}^{ss}|^2 \beta_{1}^{ss}}.
\end{equation}

%We can also substitute $\beta^{ss}_{1}=\beta^{ss}_{1,r}+i\beta^{ss}_{1,i}$ and obtain
%Eq. \ref{Ueff} 

%\begin{equation} \label{Ueff}
%U_{\rm eff} = \frac{\Gamma \left(S_{y} \delta\omega_{q_1} - \Gamma \beta^{ss}_{1,r}\right)}{2 S_{y}^{2} \Omega_L} - \frac{i \Gamma^{2} \left(S_{y} - \beta^{ss}_{1,i}\right)}{2 S_{y}^{2} \Omega_L}.
%\end{equation} 
%where $\beta_{1,r}^{ss}$ and $\beta_{1,i}^{ss}$ are the real and imaginary parts of $\beta_{1}^{ss}$ respectively given by 
%\begin{equation}  \label{betassri}
%\beta^{ss}_{1,r} = \frac{|\beta^{ss}_{1}| S_x}{\sqrt{S_x^2+S_y^2}} \qquad \beta^{ss}_{1,i} = \frac{|\beta^{ss}_{1}| S_y}{\sqrt{S_x^2+S_y^2}},
%\end{equation}
%and
%\begin{widetext}
%\begin{equation}
%S_x = - \frac{\Omega_L \left((\delta\omega_{q_1}+2U)^{2} \delta\omega_{q_1} + 12 U \Omega_L^{2} + 9 \Gamma^{2} \delta\omega_{q_1}\right)}{(\delta\omega_{q_1}+2U)^{2} \left(\Gamma^{2} + \delta\omega_{q_1}^{2}\right) + 4 (\delta\omega_{q_1}+2U) U \Omega_L^{2} + 12 U \Omega_L^{2} \delta\omega_{q_1} + 9 \Gamma^{4} + 9 \Gamma^{2} \delta\omega_{q_1}^{2}} \label{S_x}
%\end{equation}
%\begin{equation}
%S_y = - \frac{\Gamma \Omega_L \left((\delta\omega_{q_1}+2U)^{2} + 9 \Gamma^{2}\right)}{(\delta\omega_{q_1}+2U)^{2} \left(\Gamma^{2} + \delta\omega_{q_1}^{2}\right) + 4 (\delta\omega_{q_1}+2U) U \Omega_L^{2} + 12 U \Omega_L^{2} \delta\omega_{q_1} + 9 \Gamma^{4} + 9 \Gamma^{2} \delta\omega_{q_1}^{2}}.  \label{S_y}
%\end{equation}
%\end{widetext}
%where $\Gamma = \Gamma_L + \Gamma_R$ and $|\beta_{1}^{ss}| = \sqrt{(\beta_{1,r}^{ss})^2 + (\beta_{1,i}^{ss})^2 }$. 

For a single qubit, we find almost perfect agreement between THLE and the modified QCA from its inception, apart from small deviations due to truncation error. Quite remarkably, we find that the modified equations work well even for longer chains. The advantage of using quasi-classical equations lies in its simplicity and computational efficiency, which enables us to study the transmission for much longer system sizes which are computationally more challenging. 
 
 The correction for the on-site interaction can also be carried out for the more realistic side-coupled system. Again, for the side-coupled system, we try to replicate the THLE results using the QCA and obtain a similar expression for $U_{\rm eff}$ (see Appendix~\ref{SideUeffDerivation} for details). For a single qubit-resonator pair, the complex on-site interaction also nicely captures the inelastic scattering of the photons and related reduction in transmission.

As $U_{\rm eff}$ was calculated using the single site case for both the media, it does not depend on $J_x$. Nevertheless, we find a good agreement between the THLE and modified QCA for greater system sizes as long as relatively low values of $J_x~(<\Gamma_L=\Gamma_R)$ are used. However, an increase in $J_x$ causes disagreements between the THLE and modified QCA. Albeit, the corrected QCA gives quite useful results for a good range of parameters in both the media.

For the direct-coupled qubit array, within the limited system sizes that we could explore using the THLE analysis, we find very little system size dependence of the resonant transmission. This suggests that the inelastic scattering of the photons and photon transmission reduction caused by on-site interactions seem to be dominated at the two ends. This is also consistent with earlier studies on two level atoms in absence of nearest neighbor interactions~\cite{Manasi2018}. However, we end up causing too much  of inelastic scattering of the photons and transmission reduction using a homogeneous complex on-site interaction everywhere in the modified QCA. In fact, the resonant transmission for a homogeneous complex on-site interaction decreases with increasing system size and eventually falls to zero. This mismatch can be surmounted by applying complex on-site interaction only at the two ends of the medium while using its real part in the bulk of the chain. With this fix, we obtain system size independent transmission upto a chain of 300 qubits.

\subsection{Transmission probability for the medium}
To obtain the output transmission from the right end of the medium, we calculate the solutions of the coupled differential equations in the THLE analysis for the medium. The intensity of incoming light from the left side of the medium is $I_{\rm in}$~$(=E_p^2/2\pi v_g^2)$. For the direct-coupled qubit medium, the transmission probability, at time $t$, for a medium of $N$ qubits is given as
\begin{equation} \label{E_trans}
    \mathcal{T}_{\rm THLE}(t) = \frac{2\Gamma_R}{v_g I_{\rm in}} \langle b_N^{\dagger} b_N \rangle,
\end{equation}
where $\langle b_N^{\dagger} b_N \rangle$ is the number operator expectation value at the output qubit. We can similarly obtain the QCA transmission probability by calculating the expectation value of the qubit operator $\beta_N$ from the QCA. The QCA transmission probability is given as
\begin{equation} \label{QC_trans}
    \mathcal{T}_{\rm QCA}(t) = \frac{2\Gamma_R}{v_g I_{\rm in}}\beta_N^{*}(t)\beta_N(t).
\end{equation}

For the side-coupled qubits, the output photons are coming out from the resonator at the rightmost end and not from the qubit as for the direct-coupled qubit system. Hence, for a medium of $N$ qubit-resonators pairs, the transmission probability is given as
\begin{equation} \label{Tesc}
    \mathcal{T}_{\rm THLE}(t) = \frac{2\Gamma_R}{v_g I_{\rm in}} \langle f_N^{\dagger} f_N \rangle,
\end{equation}
where $\langle f_N^{\dagger} f_N \rangle$ is the expectation value of the number operator of the rightmost resonator. It is found by substituting the steady-state solutions of the qubits' variables in the integrated output resonator's equation at steady-state. For example, for a single qubit-resonator pair, to find the resonator number operator ~($F_{11}$), we multiply Eq.~\ref{longTF01} with its conjugate, to obtain the final relation as
\begin{equation} \label{F11final}
    F_{11} = \frac{\Omega_L^2 + g_{1}\Omega_L(\mathcal{S}_{01} + \mathcal{S}_{10}) + g_{1}^2\mathcal{S}_{11}}{AA^*}.
\end{equation}
 Longer system sizes can also be dealt in a similar manner. Similar to Eq.~\ref{QC_trans}, we can obtain the transmission probability for the side-coupled array of qubits using the quasi-classical equations as
\begin{equation} \label{Tqcsc}
    \mathcal{T}_{\rm QCA}(t) = \frac{2\Gamma_R}{v_g I_{\rm in}}\alpha_N^{*}(t)\alpha_N(t).
\end{equation}

The reflection probability in the THLE is given by
\begin{equation}
    \mathcal{R}_{\rm THLE}(t)= 1 + \frac{2\Omega_L}{v_g I_{\rm in}}Im[\langle\chi_1\rangle] + \frac{2\Gamma_L}{v_g I_{\rm in}}\langle \chi_1^{\dagger}\chi_1\rangle,
\end{equation}
where $\chi_1$ is $b_1$ for the direct-coupled qubits medium and is $f_1$ for the side-coupled qubits medium. We have explicitly verified numerically that $\mathcal{T}_{\rm THLE}^{\infty}+\mathcal{R}_{\rm THLE}^{\infty}=1$ (Fig. \ref{app3}).

\section{Results}

%We now present detailed results for both the models.
%for the side-coupled system and compare to the direct-coupled system. 
%The output transmission from the right end of the medium are given by (see appendix)
%\begin{equation} \label{E_trans}
    %\mathcal{T}_{\rm THLE}(t) = \frac{2\Gamma_R}{v_g I_{\rm in}} \langle b_N^{\dagger} b_N \rangle,
    %\mathcal{T}_{\rm}(t) = \frac{2\Gamma_R}{v_g I_{\rm in}} \langle n_N (t)\rangle,
%\end{equation}
%\begin{equation} \label{QC_trans}
    %\mathcal{T}_{\rm QCA}(t) = \frac{2\Gamma_R}{v_g I_{\rm in}}\beta_N^{*}(t)\beta_N(t).
%\end{equation}
%where $I_{\rm in}=E_p^2/2\pi v_g^2  , \langle n_N(t) \rangle= \langle b_N^{\dagger} b_N \rangle (\langle f_N^{\dagger} f_N \rangle) $ for direct (side) coupled THLE and $|\beta_N(t)|^2$ $(|\alpha_N(t)|^2)$ for direct (side) coupled QCA and MQCA systems.

\subsection{Direct-coupled array of qubits}
For the direct-coupled array of $N$ qubits, we calculate the steady-state solutions using THLE, QCA and modified QCA. We first set $v_g=1$ and measure various parameters in units of qubit frequency $\omega_a$. We assume all qubits in their ground state before irradiation with the photons, i.e. $\mathbf{\mathcal{S}}(t \rightarrow t_0) = 0$ and $\beta_j(t \rightarrow t_0)=0$. We use the steady-state solutions of Eqs.~\ref{mateq} and \ref{longTF01} to obtain the steady-state transmission probabilities: $\mathcal{T}^{\infty}_{\rm THLE}$ and $\mathcal{T}_{\rm QCA}^{\infty}$. For the modified QCA we replace the on-site interaction $U$ with $U_{\rm eff}$ and obtain the transmission probability $\mathcal{T}_{\rm MQCA}^{\infty}$.

\begin{figure}
    \centering
    \includegraphics[width=\linewidth]{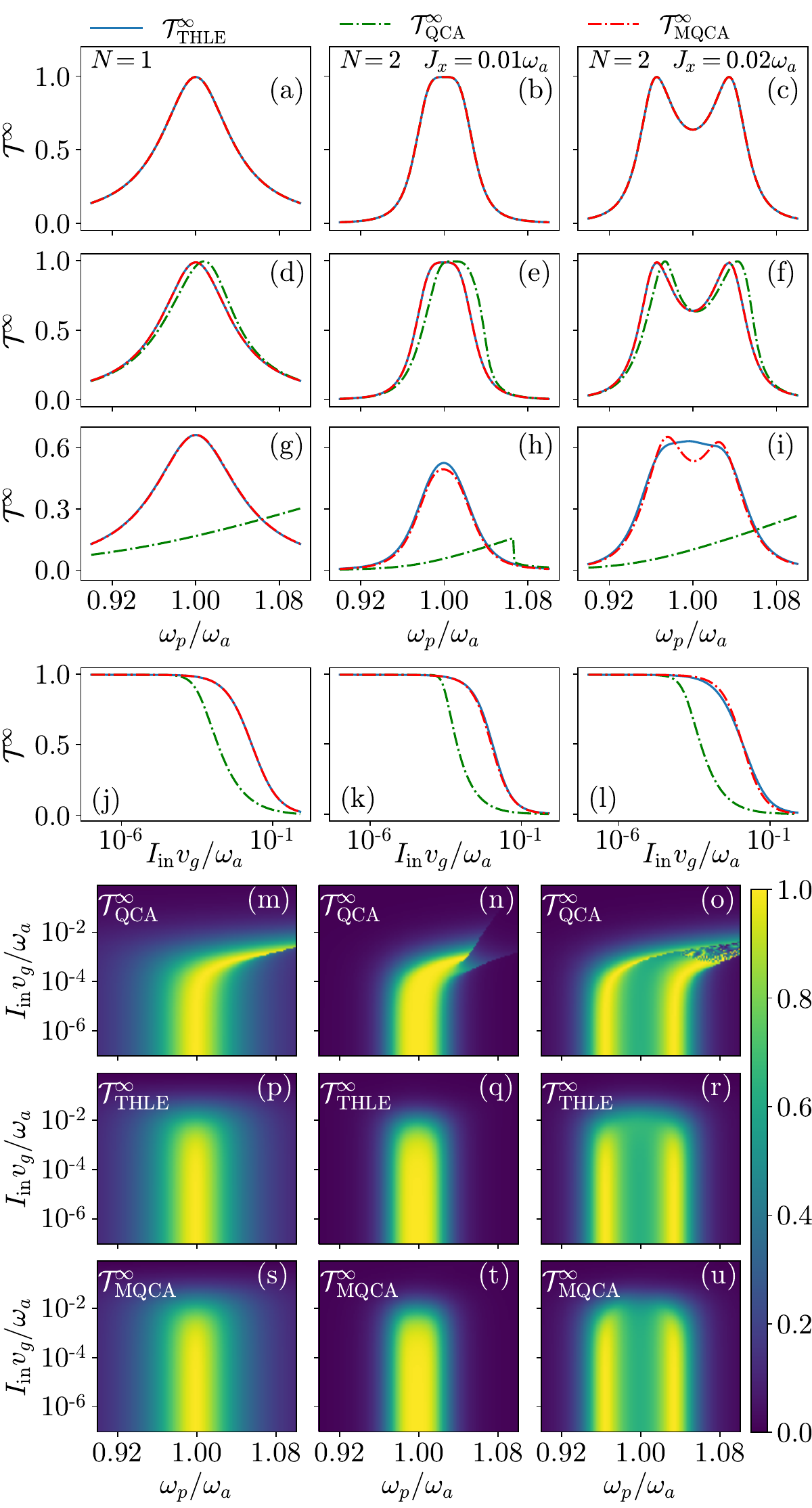}
    \caption{Single and two direct-coupled qubits steady-state transmission results comparing between the THLE~($\mathcal{T}_{\rm THLE}^{\infty}$), quasi-classical~($\mathcal{T}_{\rm QCA}^{\infty}$) and modified QCA~($\mathcal{T}_{\rm MQCA}^{\infty}$) analyses. First column shows transmission results for a single qubit while the other two columns show transmission results for two direct-coupled qubits at different values of $J_x=0.01 \omega_a$ and $J_x=0.02 \omega_a$. The first three rows show the effect of increasing the input intensity~$I_{\rm in}$ (in terms of  $\omega_a/v_g$) of photons as $I_{\rm in}=1.12 \times 10^{-6}$(a,b,c), $I_{\rm in}=1.5 \times 10^{-4}$(d,e,f) and $I_{\rm in} = 0.01$(g,h,i). The fourth row shows transmission with increasing $I_{\rm in}$ at $\omega_p= \omega_a$(j,k) and $\omega_p=0.965 \omega_a$(l). The last three rows show variation in transmission profile for a range of $I_{\rm in}$ for the QCA, THLE and modified QCA analyses respectively. Other parameters are: $\omega_{q_1}=\omega_{q_2}= \omega_a$, $\Gamma_L=\Gamma_R=0.02 \omega_a$, $U=1.05 \omega_a$. }
    \label{fig:Fig1}
  
\end{figure}

We show in Fig.~\ref{fig:Fig1} the steady-state transmission for a single qubit and a pair of two directly coupled qubits using THLE, QCA and modified QCA analyses. For both single qubit and two direct-coupled qubit mediums, all the transmission results are calculated with maximum six bosons per qubit~($m=6$) for the THLE analysis. For a single qubit, we show the transmission profiles, i.e. $\mathcal{T}_{\rm THLE}^{\infty}$, $\mathcal{T}_{\rm QCA}^{\infty}$ and $\mathcal{T}_{\rm MQCA}^{\infty}$ versus scaled frequency $\omega_p/\omega_a$ for different input intensities~($I_{\rm in}$) in Fig.~\ref{fig:Fig1}(a,d,g).
At low input intensity, 
%Fig.~\ref{fig:Fig1}(a), there is perfect resonant photon transport through the medium and we observe that the transmission profile for a single qubit has a single peak with maximum possible transmission at resonance. The
the three analyses, agree completely with each other.
%However, as we increase the intensity, Fig.~\ref{fig:Fig1}(d,g), the on-site interaction~($U$) at the qubits creates effective interactions between photons mediated by the qubits' excitations. Such effective interactions or optical nonlinearities cause photon blockade resulting in reduction of transport of photons, which is observed as decreasing of maximum transmission in the THLE transmission profile. Instead of showing a decrease in maximum transmission, the QCA only shows a shift in the transmission profile Fig.~\ref{fig:Fig1}(d,g), while the modified QCA agrees with the THLE results. 

The decrease in maximum transmission due to photon blockade, at resonant input frequency $\omega_p=\omega_a$, is more clear in Fig.~\ref{fig:Fig1}(j), where we show lowering of transmission with increasing intensity. The reduction of transmission at very high light intensities is due to the saturation of qubits by photons, which completely blocks any transmission of light to the other end. This results in no interaction between the remaining photons. The observed decrease in the QCA transmission is due to the shifting of the transmission profile, and as mentioned before there is always some frequency with perfect transmission in this analysis, Fig.~\ref{fig:Fig1}(m-o). We show the systemic variation of the entire transmission profile with intensities in  Fig.~\ref{fig:Fig1}(m,p,s) for QCA, THLE and modified QCA respectively. We observe that in all the single qubit transmission results, the QCA fails to match the THLE analysis, while the modified QCA has perfect agreement with THLE results for a large range of $I_{\rm in}$.
%This is not surprising as the $U_{\rm eff}$ was derived from the THLE of the single qubit model itself, and just signifies that the truncation errors are not significant.
In the second and third columns of Fig.~\ref{fig:Fig1}, we present the results for two directly coupled qubits at different inter qubit coupling strengths~$(J_x)$. In the second column, which has low inter qubit coupling strength ($2J_x=\Gamma_L=\Gamma_R$), we observe a single peak in the low intensity transmission profile, which is, in-fact due to the overlapping of the two peaks appearing at resonance for the two qubits in the system, Fig.~\ref{fig:Fig1}(b). In the last column, we increase $J_x$ ($J_x=\Gamma_L=\Gamma_R$), which shifts the resonance peaks away from each other and we observe the two peaks separately, Fig.~\ref{fig:Fig1}(c).
%Again, both QCA and modified QCA analyses are able to capture all the details of THLE analysis at low intensity with perfection. Similar to the single qubit case, a higher intensity causes the photon blockade phenomenon resulting in a decreased photon transport in the medium, Fig.~\ref{fig:Fig1}(e,f,h,i). In contrast to the single qubit case, photon blockade at high $I_{\rm in}$ causes the transmission profile features in the THLE analysis for the two qubit medium to change, Fig.~\ref{fig:Fig1}(i), as the two peaks appearing at low intensities have now disappeared. We observe that THLE shows strong disagreements with QCA, while a reasonable agreement with modified QCA at high intensities.
Again, for both QCA and modified QCA at low intensity two peaks appear but at high intensity they disappear.

We want to stress that $U_{\rm eff}$ is calculated by considering only single qubit and no additional fitting has been done, and still the modified QCA has captured the transmission reduction due to photon blockade fairly accurately. However, there is larger disagreement between these two analyses as $J_x$ is increased. Even with a few limitations, the modified QCA is able to capture the photon blockade effects predicted by the THLE analysis reasonably well for a large set of parameters. Transmission from the three analyses as function of $I_{\rm in}$ for two qubit model at different $J_x$ is shown in Fig.~\ref{fig:Fig1}(k,l), where the transmission decreases with increasing $I_{\rm in}$. We see a slightly larger disagreement between THLE and modified QCA transmission at higher $J_x$. The overall variation in transmission profile of the QCA, THLE and modified QCA with intensity can be seen by comparing the last three rows of Fig.~\ref{fig:Fig1}, which shows the shift in the QCA transmission similar to the single qubit and the effectiveness of the modified QCA in matching the THLE transmission results.

\begin{figure}
    \centering
    \includegraphics[width=\linewidth]{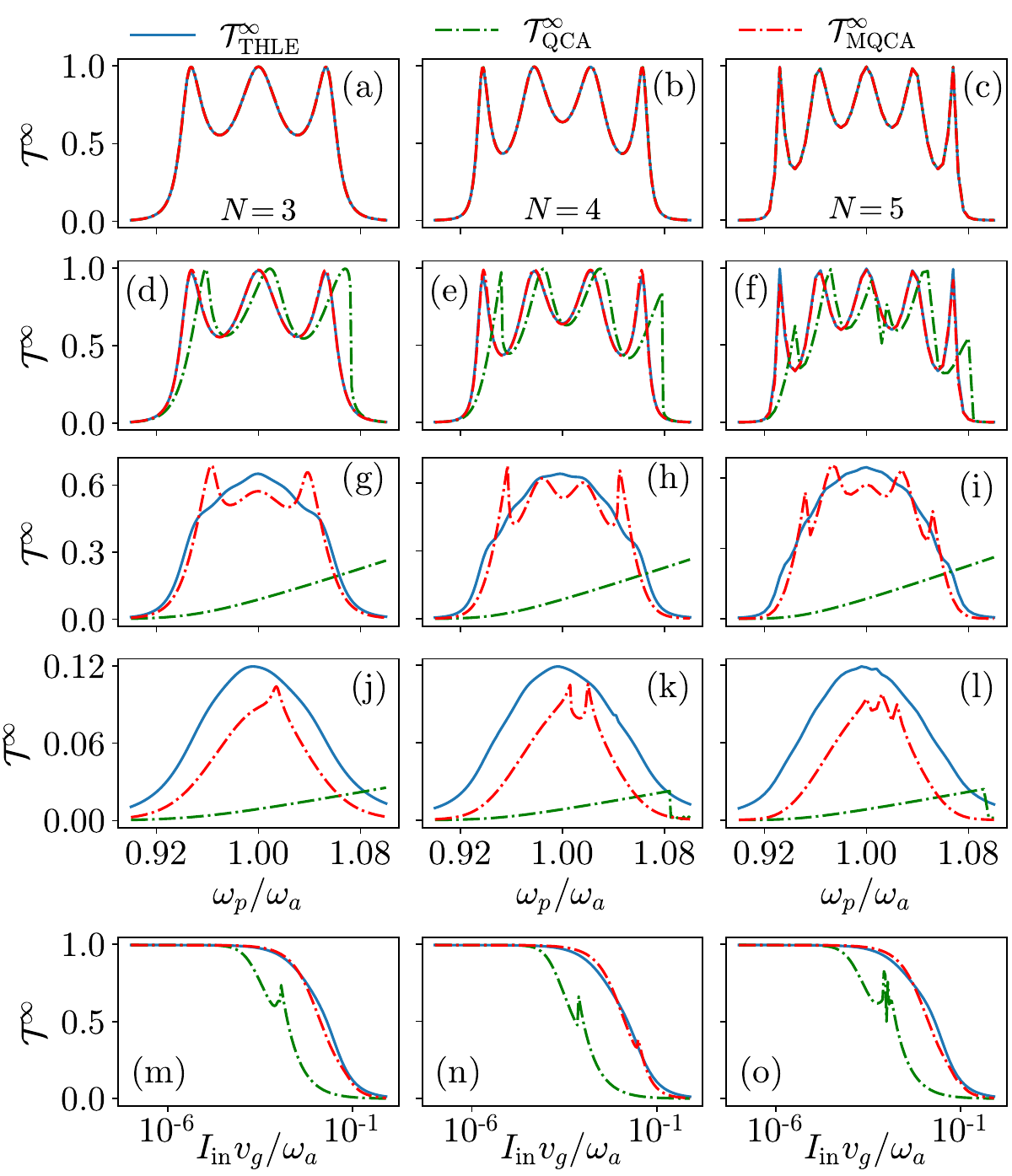}
    \caption{Longer direct-coupled system steady-state transmission for the THLE~$(\mathcal{T}^{\infty}_{\rm THLE})$, quasi-classical~$(\mathcal{T}^{\infty}_{\rm QCA})$ and modified QCA~$(\mathcal{T}^{\infty}_{\rm MQCA})$ analyses. The three columns show transmission profiles for system sizes ($N$) of 3, 4, and 5 qubits respectively. In the first four rows we show transmission profiles with input intensity $I_{\rm in}$(in terms of $\omega_a /v_g$) increasing on going down the columns as  $I_{\rm in} = 1.12 \times 10^{-6}$(a,b,c), $I_{\rm in} = 1.5 \times 10^{-4}$(d,e,f), $I_{\rm in} = 0.01$(g,h,i) and $I_{\rm in} = 0.1$(j,k,l). The last row shows transmission with intensity at input frequency~$\omega_p=\omega_a$ for $N=3,5$(m,o) and $\omega_p=0.978\omega_a$ for $N=4$(n). The remaining parameters are, $\omega_{q_j}=\omega_a$, $\Gamma_L=\Gamma_R=0.02\omega_a$, $J_x=0.02\omega_a$, $U=1.05\omega_a$.}
    \label{fig:Fig2}
\end{figure}

We further show a comparison of the transmission profiles for the three analyses for upto five directly coupled qubits in Fig.~\ref{fig:Fig2}. The three columns in Fig.~\ref{fig:Fig2} correspond to increasing system sizes of three, four and five qubits for the direct-coupled qubits medium. In the THLE analysis, the three, four and five direct-coupled qubits array have truncations at $m=5,3$ and $2$ respectively. The truncation is larger, i.e. $m$ decreases, for larger system sizes due to the computational capacity being reached. However, due to large on-site interaction $U$, we expect the truncation errors to be negligible though we cannot check this explicitly for the largest system size. In the first four rows, we show the transmission profile with increasing $I_{\rm in}$ down the column. At low $I_{\rm in}$, Fig.~\ref{fig:Fig2}(a,b,c), we observe perfect resonant transmission with the number of peaks in the THLE transmission profile increase as resonance points increases with the system size. As the $I_{\rm in}$ is increased, similar to the single and two direct-coupled qubits case in Fig.~\ref{fig:Fig1}, the THLE transmission profile shows the effect of photon blockade as maximum transmission decreases with $I_{\rm in}$, Fig.~\ref{fig:Fig2}(d,e,f). On further increase in $I_{\rm in}$, the transmission further decreases and the peaks disappear in the THLE analysis, Fig.~\ref{fig:Fig2}(g,h,i). 

\subsection{Side-coupled array of qubits}

For THLE we solve the system of qubits equations, with resonators equation integrated out, to obtain the transmission from the right end of the medium. We assume all the qubits and resonators to be in their ground state and set $v_g=1$. Unlike the direct-coupled case, we observe a reduction of photon transport due to effective photon-photon interactions only at intermediate intensities for the side-coupled system. At very large intensities, the side-coupled system starts to behave more like a resonator chain without any qubits. This is because the qubits get saturated by the photons and are no longer able to create effective photon-photon interactions blocking the photon transport through the resonator, which results in recovery of photon transport at a high intensity. %Thus the side-coupled system shows a stark difference in light transmission at a high intensity in comparison to the direct-coupled lattice investigated above. 

\begin{figure}
    \centering
    \includegraphics[width=\linewidth]{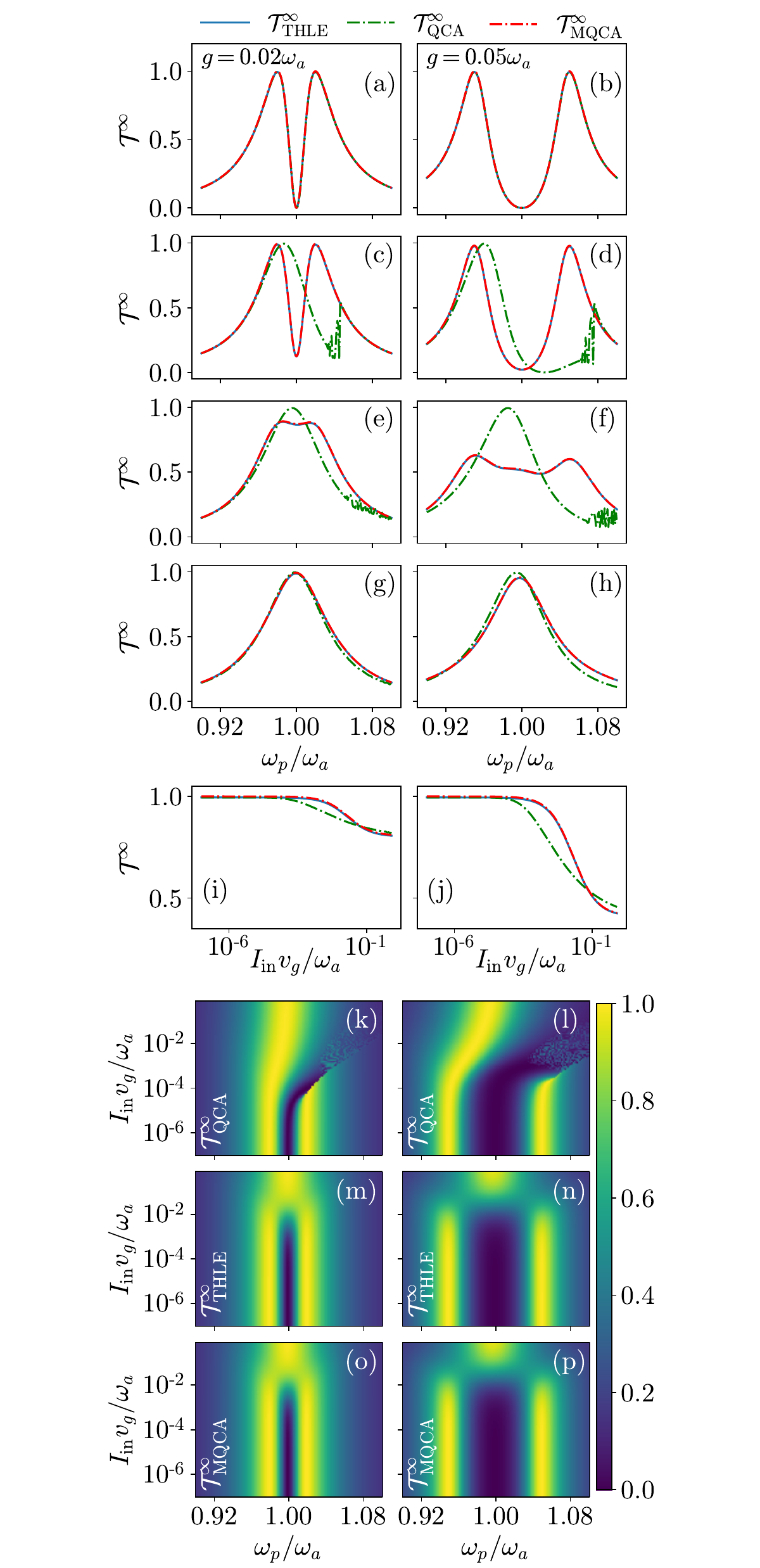}
    \caption{Single qubit side-coupled to a single resonator transmission results showing comparison between the THLE~$(\mathcal{T}^{\infty}_{\rm THLE})$, quasi-classical~$(\mathcal{T}^{\infty}_{\rm QCA})$ and modified QCA~$(\mathcal{T}^{\infty}_{\rm MQCA})$ analyses. The first four rows show the effect of increasing input intensity $I_{\rm in}$ (in terms of $\omega_a / v_g$) as $I_{\rm in} = 1.12 \times 10^{-6}$(a,b), $I_{\rm in} = 7.1 \times 10^{-4}$(c,d), $I_{\rm in} = 0.034$(e,f) and $I_{\rm in} = 0.68$(g,h) at two different resonator-qubit coupling strengths of $g_{1}=0.02\omega_a$ and $g_{1}=0.05\omega_a$ in the two columns respectively. Change in transmission on increasing $I_{\rm in}$ is shown in (i,j) at, $\omega_p=0.98\omega_a$(i) and $\omega_p=0.95\omega_a$(j). Variation of the entire transmission profile with increasing $I_{\rm in}$ is shown in the last three rows for QCA, THLE and modified QCA respectively. The other parameters are: $\omega_{r_1}=\omega_{q_1}=\omega_a$, $\Gamma_L=\Gamma_R=0.02\omega_a $ and $U=1.05\omega_a$.}
    \label{fig:Fig3}
\end{figure}

We show the transmission results for the medium of one qubit-resonator pair coupled to the baths in Fig.~\ref{fig:Fig3}. The two columns in the figure correspond to two different qubit-resonator coupling strengths~($g_{1}$), which are $g_{1}=0.02\omega_a$ on the left and $g_{1}=0.05\omega_a$ on the right. The basic features of the side-coupled qubits' transmission profile at low intensity can be seen in Fig.~\ref{fig:Fig3}(a,b), where we observe two transmission peaks with perfect resonant photon transport separated by a large valley with a minima in transmission at $\omega_p=\omega_a$. We have chosen the resonator~$(\omega_{r_1})$ and qubit~$(\omega_{q_1})$ frequencies to be equal in Fig.~\ref{fig:Fig3}, however, when resonator and qubit frequencies are not equal, the minima in transmission occurs whenever $\omega_p$ is equal to the qubit frequency. The maxima in transmission at low intensities of incident light are near the resonance frequencies (or normal modes) of the isolated medium, and the minima in between the two resonance frequencies is due to destructive interference (see Appendix~\ref{app_rv1}). %The minima in transmission is seen when the qubit absorbs an incoming photon and scatters the other incoming photons while it is excited, effectively creating a nonlinear interaction among the photons, resulting in a reduced photon transport.
 The effect of increasing $g_{1}$, at low intensity, can be seen in Fig.~\ref{fig:Fig3}(b) where the two peaks have now moved further apart and the dip between them has become broader, which suggests an increased influence of the qubit in the medium.
 %At low intensity, both QCA and modified QCA agree to the THLE results perfectly as in the case of direct-coupled qubits.

In the first four rows, $I_{\rm in}$ gradually increases on going down the columns. The THLE analysis shows that the increase in intensity initially causes disruption of photon transport and the transmission peaks to lower down, Fig.~\ref{fig:Fig3}(c)-(h), which is similar to lowering of transmission observed in the direct-coupled qubits system. The reduction in transmission is again due to the effective photon-photon interactions at higher intensities mediated by the on-site interaction present in the qubits. Such effective interactions block the propagation of light. We observe that higher values of $g_{1}$ in the second column causes a greater reduction of transmission as the influence of the qubits on generating the effective photon-photon interactions has increased, Fig.~\ref{fig:Fig3}(e,f). We see that with increase in $I_{\rm in}$, the valley between the peaks rises gradually until it supersedes the lowering of transmission. At this moment, the peaks disappear and we observe maximum photon blockade and reduction of transmission. At even higher intensity, Fig.~\ref{fig:Fig3}(g,h), the rise of transmission at $\omega_p=\omega_a$ continues and the transmission profile starts to look a bit like the one for a single resonator. This occurs as the qubit gets saturated by photons and most incoming photons pass through the resonator and transmitted with very little influence of the qubit. Thus, there are no longer effective photon-photon interactions and related photon blockade in transmission at very high intensities.  

%In the Fig.~\ref{fig:Fig3}(c)-(f), we also observe that the QCA fails to match the transmission profile with increasing intensity. We observe an overall shift in the QCA transmission profile where the right half of the profile shows shifting as well as lowering of transmission and the left part moves towards the center. Eventually at high intensity, the QCA transmission profile becomes similar to that of the resonator as the qubit saturates. The modified QCA, however, shows perfect agreement with THLE results for all $I_{\rm in}$ values which is not surprising as the correction ($U_{\rm eff}$) was derived from the THLE for single resonator-qubit pair with enough truncation to capture the photon blockade and saturation phenomena.  

We show the variation of transmission with intensity in Fig.~\ref{fig:Fig3}(i,j). We see that the QCA has matched the photon blockade of transmission as predicted by the THLE to some extent, but the modified QCA results have matched the THLE results perfectly. We must note that any lowering of transmission in QCA is due to shifting of transmission profile and not due to an effective photon-photon interactions. These results shows that the QCA is accurate for some parameters in the side-coupled system but not for all parameters, in particular when the transmission is low. On the other hand, for all parameters, the modified QCA has shown perfect agreement. This is further confirmed by the variation of the transmission profile with the intensity as shown in the last three rows of Fig.~\ref{fig:Fig3} for QCA, THLE and modified QCA respectively. We observe that the QCA transmission profile is shifting towards right with increasing intensity, Fig.~\ref{fig:Fig3}(k,l) and shows little agreement with the THLE results at intermediate intensities. The modified QCA, however, has perfectly captured all the features predicted by the THLE analysis, Fig~\ref{fig:Fig3}(m,n,o,p).
 %At $\omega_p=1.0$, we see the rising of the transmission minima between the peaks with increasing $I_{\rm in}$ and here we see the QCA failing to capture the phenomenon while the modified QCA again matches it well.

\begin{figure}
    \centering
    \includegraphics[width=\linewidth]{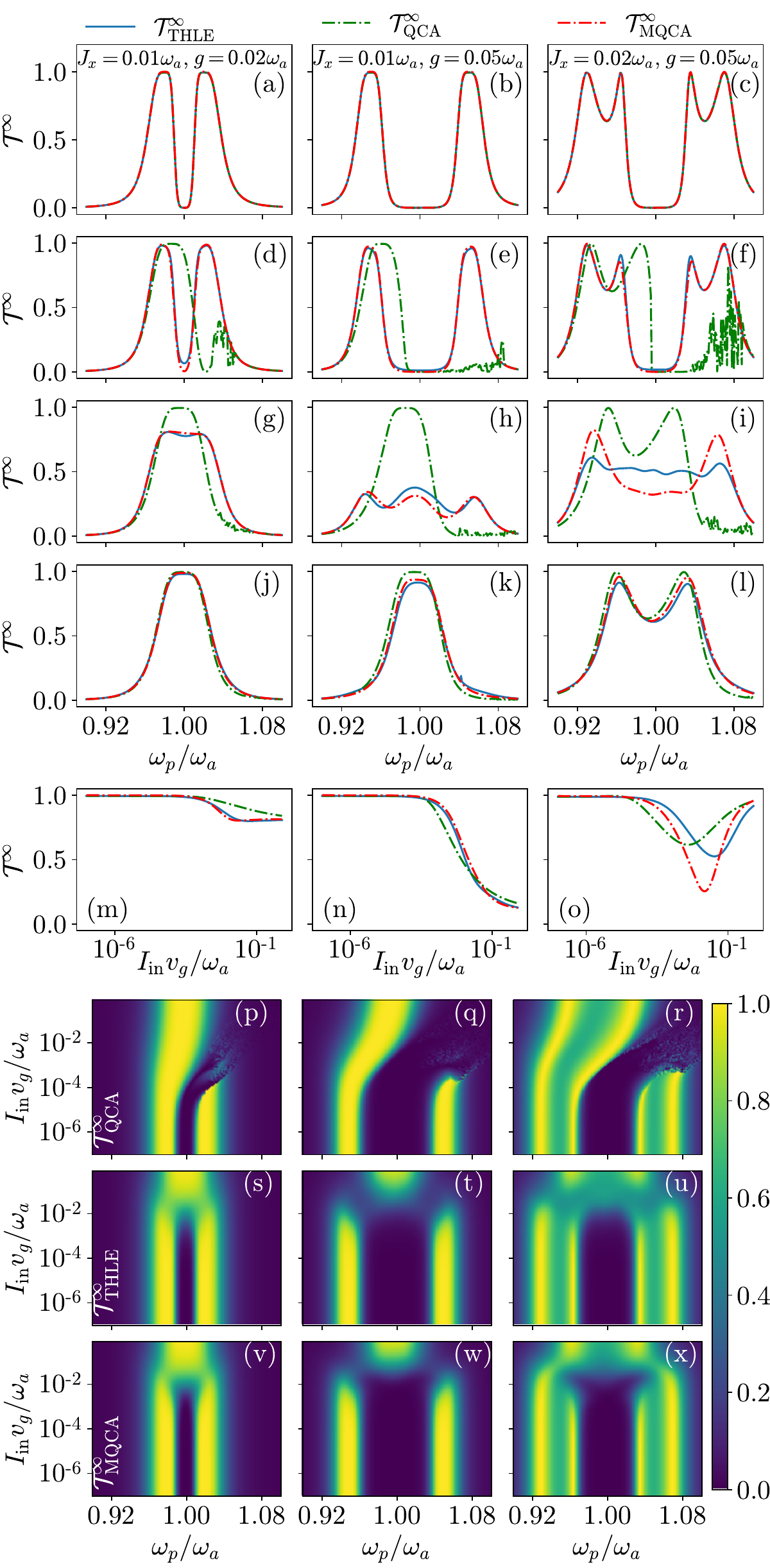}
    \caption{Two side-coupled qubits transmission results showing the effects of inter-resonator coupling $(J_x)$ as well as a comparison between the THLE~$(\mathcal{T}^{\infty}_{\rm THLE})$, quasi-classical~$(\mathcal{T}^{\infty}_{\rm QCA})$ and modified QCA~$(\mathcal{T}^{\infty}_{\rm MQCA})$ analyses. The three columns have $(J_x / \omega_a,g_{j} / \omega_a)=(0.01,0.02), (0.01,0.05)$ and $(0.02,0.05)$ respectively. The first four rows show the effect of increasing input intensity $I_{\rm in}$ (in terms of $\omega_a / v_g$) as $I_{\rm in} = 1.12 \times 10^{-6}$(a,b,c), $I_{\rm in} = 7.1 \times 10^{-4}$(d,e,f), $I_{\rm in} = 0.034$(g,h,i) and $I_{\rm in} = 0.68$(j,k,l). Transmission with increasing $I_{\rm in}$ is shown in (m,n,o) at $\omega_p=0.98\omega_a$(m), $\omega_p=0.95\omega_a$(n) and $\omega_p=0.965\omega_a$(o). Variation of the entire transmission profile with increasing $I_{\rm in}$ is shown in the last three rows each for QCA, THLE and modified QCA respectively. The other parameters are: $\omega_{r_j}=\omega_{q_j}=\omega_a$, $\Gamma_L=\Gamma_R=0.02\omega_a $ and $U=1.05\omega_a$.}
    \label{fig:Fig4}
\end{figure}

We show the transmission results for a medium of two qubits coupled to a chain of two resonators at different inter-resonator coupling strengths~$(J_x)$ and resonator-qubit coupling~$g_{j}$ in Fig.~\ref{fig:Fig4}. Each column has constant $J_x$, where the first two columns have low $J_x(=\Gamma_L / 2 =\Gamma_R / 2)$ and the last column has high $J_x(=\Gamma_L =\Gamma_R)$. Each column also has a constant $g_{j}$ with low $g_{j}(=0.02\omega_a)$ in the first column and high $g_{j}(=0.05\omega_a)$ in the last two columns. At low intensity and $J_x$, the transmission profile from the THLE analysis shows perfect resonant photon transport with two peaks separated by a valley with no photon transport at $\omega_p=\omega_a$, Fig.~$\ref{fig:Fig4}$(a,b). Increasing $g_{j}$ causes the resonance peaks to move away from each other, Fig.~\ref{fig:Fig4}(a,b). The individual peak splits as we increase $J_x$ which results into four peaks in the transmission profile with no transmission at $\omega_p=\omega_a$, Fig.~\ref{fig:Fig4}(c). 

The input intensity~($I_{\rm in}$) is increased as we go down the column in the first four rows. In all cases, from the THLE analysis we observe disruption of photon transport with increase in $I_{\rm in}$ due to effective photon-photon interaction along with gradual increase in transmission at $\omega_p=\omega_a$. At low $J_x$, the photon-photon interaction causes lowering of transmission peaks, similar to the single resonator qubit pair case in Fig.~\ref{fig:Fig3}, where the transmission keeps decreasing till intermediate intensities, Fig.~\ref{fig:Fig4}(d,e,g,h). At even higher intensity, the qubits starts to get saturated by photons and a large photon transport is restored in the absence of inelastic scattering of the photons, Fig.~\ref{fig:Fig4}(j,k). At higher $J_x$, we observe from the THLE analysis that out of four peaks the ones near $\omega_p=\omega_a$ lose transmission first as we increase $I_{\rm in}$ as shown in Fig.~$\ref{fig:Fig4}$(f). With further increase in $I_{\rm in}$, the effective photon-photon interactions and related photon blockade increase, and their effect is seen in the entire transmission profile, Fig.~\ref{fig:Fig4}(i).
%At even higher $I_{\rm in}$, the qubits start to get saturated by photons with large photon transport getting restored which, in the case of two resonators, shows two peaks as seen in Fig.~\ref{fig:Fig4}(l). At very high intensities we essentially observe transmission through a harmonic chain of resonators with little effects from the qubits. In this limit, the qubits can be thought of as additional baths, and the photon transport is mainly by the resonator chain.

%The QCA fails to match the transmission profile of the THLE analysis with increasing $I_{\rm in}$ and is unable to capture the photon blockade phenomenon (as well as saturation of qubits) for most parameters. The QCA transmission profile instead shifts towards increasing $\omega_p$ when compared to the THLE results, Fig.~\ref{fig:Fig4}(d,e,f). On the other hand, the modified QCA captures the reduction of photon transport and other features predicted by the THLE fairly accurately at low $J_x$. With increased $J_x$ in the third column, the modified QCA is still able to capture the features of the transmission profile but shows lesser agreement with THLE analysis in predicting the reduction of transmission.

 The overall effect of $I_{\rm in}$ on transmission is shown in Fig.~\ref{fig:Fig4}(m,n,o), which shows reduction of transmission peak due to effective photon-photon interactions for the three analyses at various parameters. At low $J_x$, Fig.~\ref{fig:Fig4}(m,n), reduction of transmission predicted by THLE is captured fairly accurately by the modified QCA. The QCA does not capture the reduction of transmission accurately in Fig.~\ref{fig:Fig4}(m) but shows agreement at some intermediate intensities with the THLE analysis in Fig.~\ref{fig:Fig4}(n). We must note that reduction of transmission in QCA is due to shifting of transmission profile and not because of photon blockade. At higher $J_x$, Fig.~\ref{fig:Fig4}(o), modified QCA shows less accuracy in matching the lowering of transmission predicted by THLE analysis while simple QCA has even larger disagreement with THLE. We show the entire transmission profile with varying $I_{\rm in}$ in the last three rows of Fig.~\ref{fig:Fig4} for the QCA, THLE and modified QCA respectively. We can observe that for both values of $J_x$ the QCA fails to match the features of THLE analysis results for a large range of parameters. The modified QCA, however, is able to capture most of the transmission features as predicted by THLE analysis. The agreement between these two analysis is good at low $J_x$, however, we also observe that the modified QCA becomes less accurate as $J_x$ is increased. 

\begin{figure}
    \centering
    \includegraphics[width= \linewidth]{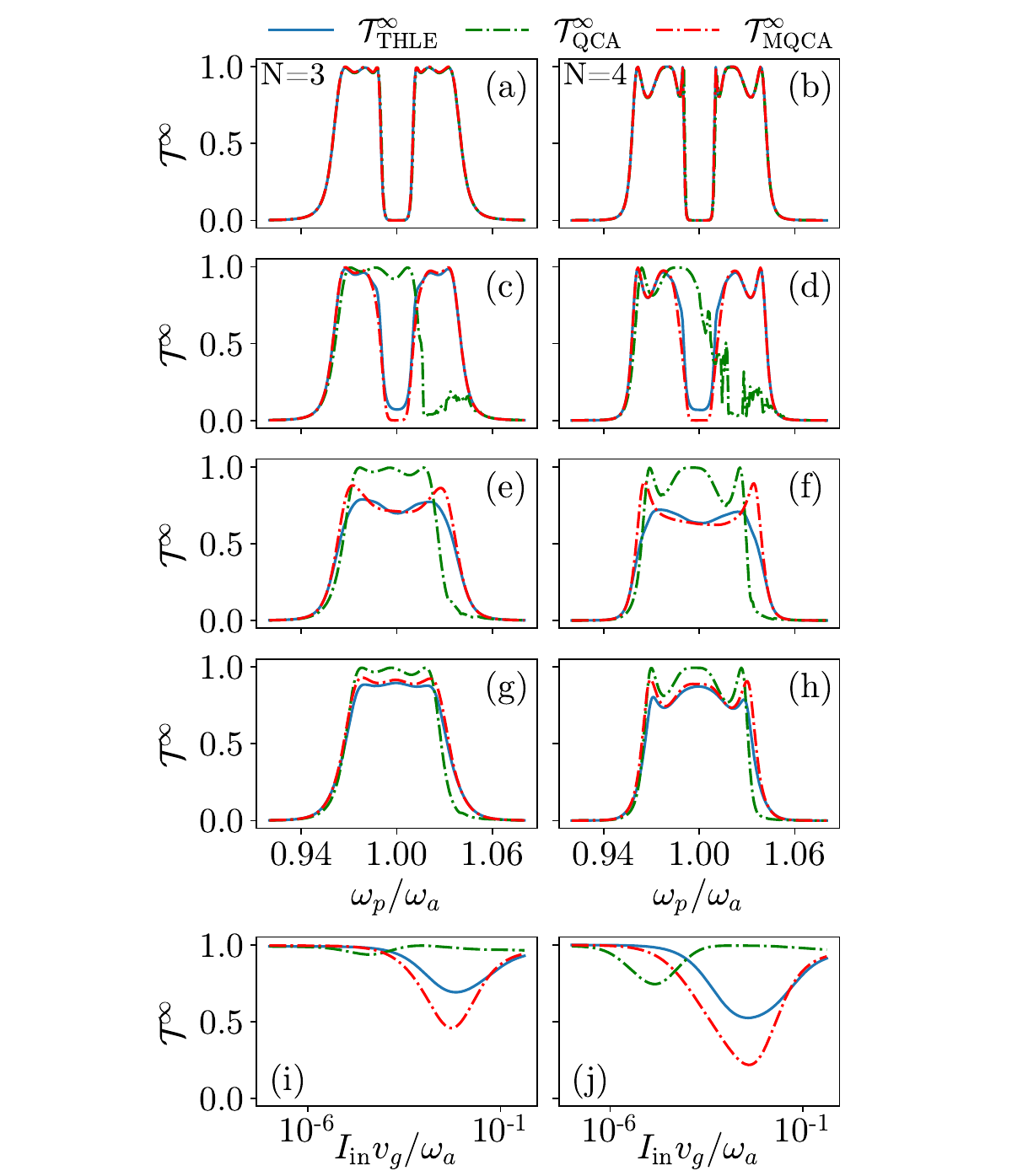}
    \caption{Longer side-coupled system steady-state transmission for the THLE~$(\mathcal{T}^{\infty}_{\rm THLE})$, quasi-classical~$(\mathcal{T}^{\infty}_{\rm QCA})$ and modified QCA~$(\mathcal{T}^{\infty}_{\rm MQCA})$ analyses. The two columns show results for system sizes ($N$) of 3 and 4 qubits respectively. The first four rows show transmission profiles with input intensity $I_{\rm in}$ (in terms of $\omega_a / v_g$) increasing on going down the columns as  $I_{\rm in} = 1.12 \times 10^{-6}$(a,b), $I_{\rm in} = 7.1 \times 10^{-4}$(c,d), $I_{\rm in} = 0.034$(e,f) and $I_{\rm in} = 0.135$(g,h). The last row shows transmission with increasing intensity for the corresponding system sizes at $\omega_p=0.9872\omega_a$(i) and $\omega_p=0.9896\omega_a$(j). The remaining parameters are, $\omega_{r_j}=\omega_{q_j}=\omega_a$, $\Gamma_L=\Gamma_R=0.02\omega_a$, $J_x=0.01\omega_a$, $g_{j}=0.02\omega_a$, $U=1.05\omega_a$.}
    \label{fig:Fig5}
\end{figure}

We show the effects of increasing system size in the side-coupled qubits medium in Fig~\ref{fig:Fig5} where the left and right columns are for three and four qubits respectively. The first four rows correspond to increasing $I_{\rm in}$ in every column. The THLE truncation is increased for two system sizes in each column with $m=4$ and $m=2$ respectively and large $U$ ensures the stability of transmission results. At low intensity, the side-coupled qubits medium shows perfect resonant photon transport with two major transmission peaks each having three smaller peaks for $N=3$ near maximum transmission, Fig.~\ref{fig:Fig5}(a). Larger system sizes have more peaks, however, in Fig.~\ref{fig:Fig5}(b) for a medium of four side-coupled qubits the middle two peaks are merged and we can only observe three peaks distinctly in each major peak. Increasing resonant peaks with system size can be observed distinctly at higher $J_x$. When we increase $I_{\rm in}$ for each system size as we go down the column, the THLE analysis shows a reduction in photon transport as well as total transmission due to the effective photon-photon interactions resulting into photon blockade at intermediate $I_{\rm in}$. The peaks closer to the center show the reduction of transmission first. At even higher intensity, Fig.~\ref{fig:Fig5}(g,h), the qubits start to get saturated by photons, which results in a revival of photon transport in the medium, and the transmission profile is primarily controlled by the features of the resonator chain. 

%Both QCA and modified QCA show perfect agreement with the THLE analysis at low $I_{\rm in}$ and capture all the features for various parameters of the model, Fig.~\ref{fig:Fig5}(a,b). However, the QCA fails to match the THLE transmission profile at intermediate $I_{\rm in}$, Fig.~\ref{fig:Fig5}(c,d), but have some agreement with THLE at very high $I_{\rm in}$ where qubit saturation has started to occur. At intermediate $I_{\rm in}$, the transmission profile of QCA shifts towards the resonant input frequency and we observe lowering of transmission peaks in the right half of the transmission profile. We would like to note here that for some parameters the QCA starts showing oscillating behavior, in which case we have shown the time average of the transmission. The modified QCA, on the other hand, continues to show good agreement as we increase the intensity till we reach intermediate intensities where the transmission reduction due to photon blockade is highest and where it also shows some deviation though not as much as QCA. At even higher $I_{\rm in}$, with saturation of qubits, the modified QCA again matches the THLE transmission profile with good accuracy.   

We show the transmission from the three analyses with increasing $I_{\rm in}$ at one of the intermediate peaks ($\omega_p=0.95\omega_a$) in the last row of Fig.~\ref{fig:Fig5}. We observe interesting non-monotonic behavior similar to Fig.~\ref{fig:Fig4}(m,o) showing a decrease in transmission due to effective photon-photon interaction and then a revival of photon transport through the resonator chain due to saturation of qubits by photons. The QCA fails to predict the reduction of transmission over a wide range of intensity, while the modified QCA has shown good agreement with THLE analysis in predicting this reduction. 

%\begin{figure}
%    \centering
%    \includegraphics[width= \linewidth]{Figure6_test.pdf}
%    \caption{$\mathcal{T}_{\rm THLE}$ curves for a medium of four resonator-qubit pairs. First and second plot from left shows transmission profile at $I_{\rm in} = 1.12 \times 10^{-6}$ and $I_{\rm in} = 0.68$ respectively. While, third figure has $\mathcal{T}_{\rm THLE}$ vs $I_{\rm in}$ at $\omega_p=1.01$. The blue curve has parameters,$g_j=0.02 and \omega_{q_j}=1.0$ and orange curve has parameters, $g_1=0.04,g_2=0.01,g_3=0.02,g_4=0.045,\omega_{q_1}=0.92,\omega_{q_2}=0.94,\omega_{q_3}=1.06,\omega_{q_4}=1.08$. The remaining parameters are, $\omega_{r_j}=1.0,\Gamma_L=\Gamma_R=0.02$, $J_x=0.01$, $U=1.05$.} 
%    \label{fig:my_label}
%\end{figure}
\subsection{Inhomogeneous side-coupled systems}
\begin{figure}
    \centering
    \includegraphics[width= \linewidth]{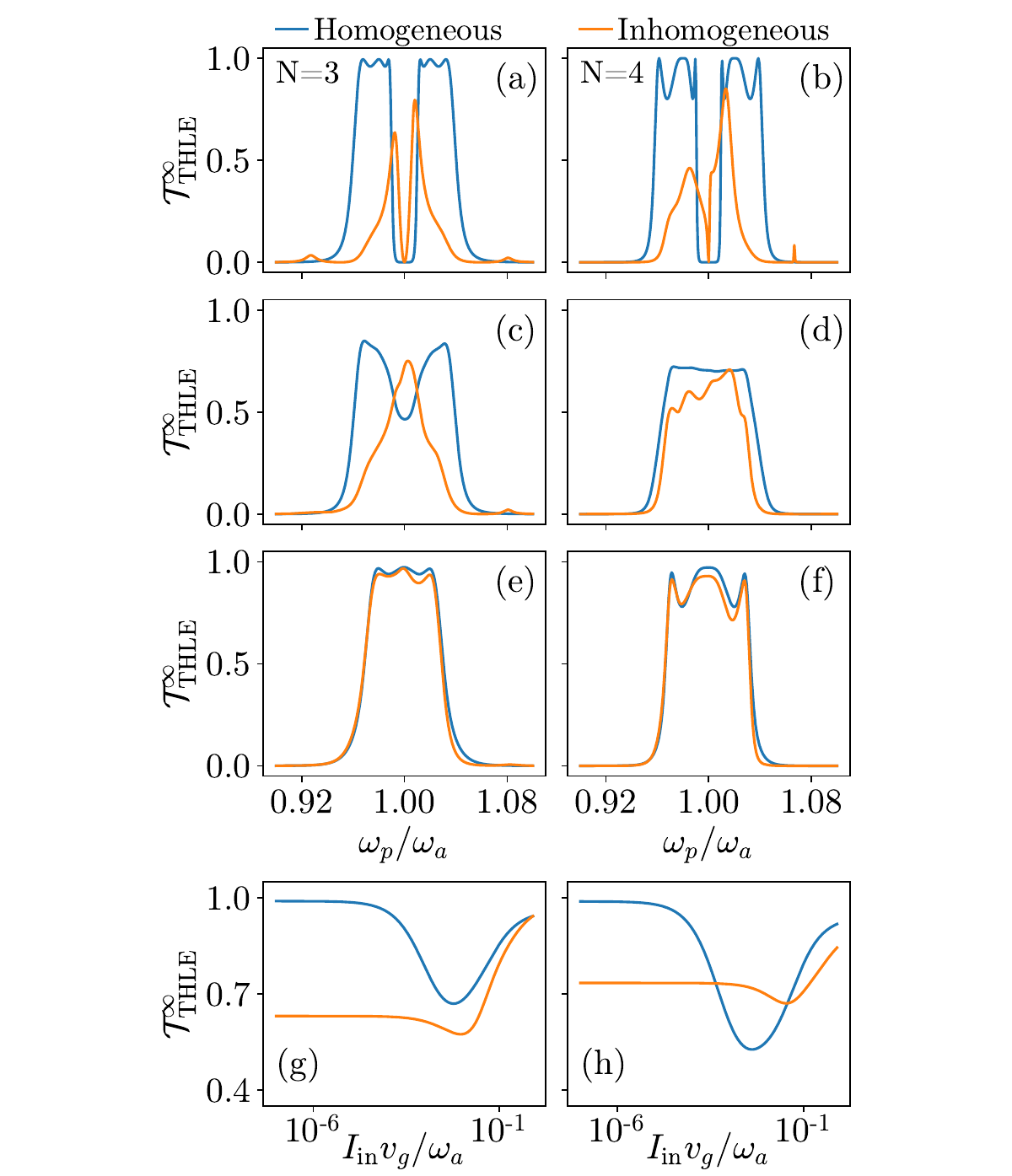}
    \caption{Inhomogeneous side-coupled systems THLE transmission profile for system sizes $N=3,4$ in the left and right columns respectively. The first three rows show transmission profiles with input intensity $I_{\rm in}$ (in terms of $\omega_a / v_g$) increasing on going down the columns as $I_{\rm in} = 1.12 \times 10^{-6}$(a,b), $I_{\rm in} = 9.4 \times 10^{-3}$(c), $I_{\rm in} = 0.047$(d) and $I_{\rm in} = 0.68$(e,f). The last row shows transmission with $I_{\rm in}$, for homogeneous and inhomogeneous qubits respectively, at $\omega_p=0.988\omega_a$ and $0.992\omega_a$ for $N=3$(g), and at $\omega_p=1.01\omega_a$ for $N=4$(h). For the homogeneous medium, both system sizes have $g_j=0.02\omega_a$ and $\omega_{q_j}=\omega_a$. For the inhomogeneous system $N=3$ has $g_j/\omega_a=(0.04,0.01,0.04)$, $\omega_{q_j}/\omega_a=(0.95,1.0,1.06)$ while $N=4$ has, $g_j/\omega_a=(0.05,0.007,0.02,0.04)$, $\omega_{q_j}/\omega_a=(0.92,1.0,1.06,1.08)$. The remaining parameters are,  $\omega_{r_j}=\omega_a$, $\Gamma_L=\Gamma_R=0.02\omega_a$, $J_x=0.01\omega_a$ and $U=1.05\omega_a$.}
    \label{fig:Fig6}
\end{figure}
Finally, in Fig.~\ref{fig:Fig6}, we consider photon transport through an inhomogeneous medium of side-coupled qubits, where we introduce inhomogeneity in qubit frequencies ($\delta\omega_{q_j}$) and the qubit-resonator couplings ($g_j$) \cite{Underwood12,See2019,Orell19}. For an inhomogeneous or disordered system, the coherent quantum behavior itself has the possibility of showing reduced transmission due to Anderson localization at low intensity as seen in Fig.~\ref{fig:Fig6}(a,b). As $I_{\rm in}$ is increased there is an initial loss of transmission due to effective photon-photon interactions even for the inhomogeneous system, \ref{fig:Fig6}(c,d). However, at high $I_{\rm in}$, photon transport re-emerges as qubits get saturated by photons and qubits have less effect on photons transiting through the resonators, Fig.~\ref{fig:Fig6}(e,f). We observe that the transmission profile for both homogeneous and inhomogeneous qubits is very similar at high $I_{\rm in}$ as the resonators in the medium are all identical. The small difference at high $I_{\rm in}$ is due to non-identical qubits being at different saturation levels. The non-monotonic reduction and rise of transmission is observed in case of both homogeneous and inhomogeneous qubits as shown in Fig.~\ref{fig:Fig6}(g,h). A very interesting effect is seen for the inhomogeneous system, namely, the effective photon-photon interactions end up resulting in a higher transmission than its photon transport at low intensity, which is lowered due to Anderson localization.

\section*{Summary and Outlook}

In this paper, we have explored transmission properties of photons through lattices of direct and side-coupled qubits, where the qubits have on-site interaction. With the increase in intensity of incoming photons in a coherent state, both lattices develop effective photon-photon interactions and related photon blockade mediated by the on-site interaction at the qubits. For the direct-coupled qubits, at single photon limit the transmission line-shape shows resonance peaks equal to the number of qubits in the system. With an increase in intensity the photon blockade occurs, which causes lowering of resonance transmission peaks until negligible transmission is observed at a very high intensity when the qubits get saturated by photons. The side-coupled qubits, at low intensity, also show resonance peaks with maximum transmission and a transmission minima at the qubit frequency. With an increase in intensity, the resonance peaks lowers due to photon blockade along with rise in transmission at the qubit frequency. At a very high intensity the qubits get saturated and their effects start to disappear resulting in a revival of photon transport in the lattice of side-coupled qubits. A non-monotonic behavior of transmission with increase in $I_{\rm in}$ is observed for both homogeneous and inhomogeneous side-coupled qubits.

We also performed a quasi-classical analysis for the two lattices, which completely fails to show any transmission reduction due to photon blockade. For both lattices, the QCA shows agreement with THLE at single particle limit but fails at higher intensities and shows a shift in the QCA transmission profile. One of our main contributions in this paper is to modify the QCA equations by introducing a complex on-site interaction by matching the THLE and QCA transmissions for the single site case. The modified QCA is effective in capturing effective photon-photon interactions for a wide range of parameters even at larger system sizes.

At low $I_{\rm in}$, both QCA and modified QCA matches the THLE transmission profile. However, as we increase the intensity, the on-site interaction ($U$) at the qubits creates effective interactions or optical nonlinearities between photons mediated by the qubits' excitations cause photon blockade resulting in reduction of transport of photons, which is observed in the THLE transmission profile. The QCA only shows a shift in the transmission profile while the modified QCA agrees with the THLE results. For the side-coupled system, at very high intensities we essentially observe transmission through a harmonic chain of resonators with little effects from the qubits. In this limit, the qubits can be thought of as additional baths, and the photon transport is mainly by the resonator chain. The modified QCA, however, shows perfect agreement with THLE results for all $I_{\rm in}$ values fairly accurately at low $J_x$. which is not surprising as the correction ($U_{\rm eff}$) was derived from the THLE for single resonator qubit pair with enough truncation to capture the photon blockade and saturation phenomena.
However, with increase in $J_x$, the effectiveness of modified QCA reduces. We hope that further studies will provide more insight into improving this approach to work for an even wider set of parameters.
%With increased Jx , the modified QCA is still able to capture the features of the transmission profile but shows lesser agreement with THLE analysis in predicting the reduction of transmission.
 
The one-dimensional open QED lattices explored in our work are complementary to an earlier experimental work in circuit QED lattices ~\cite{Fitzpatrick2017}, but we believe that the transport features observed by us are within the reach of recent experimental developments. Control over internal losses in the bulk of the lattice poses experimental challenges, and the current theoretical study can readily be extended to incorporate some intrinsic losses from the qubits and resonators. It would also be useful to examine the effects of inhomogeneity and optical nonlinearity systematically and explore the possibility of many-body localization \cite{Pal10,Roy15,Singh17, See2019, Orell19} in these systems. On the analysis side, there is room for further improvement with a more systematic way of introducing the corrections such that the modified quasi-classical equations are applicable to a much broader range of parameters. Perhaps some renormalization procedure can also be developed that considers one qubit at a time with more variables and the rest of the system as renormalized parameters. This method can also be extended using the exact result of two site system to be applicable at higher $J_x$ also.

Further, it will be of interest to see how well the modified QCA is able to match the output spectra including the phenomena such as the Mollow triplet which are observed for this system in the quantum analysis. These may lead to further insights for making better quasi-classical approximations. An earlier work has extracted an effective temperature by fitting the low frequency part of the spectra in quantum spin chains \cite{Kilda2019}. Similar fitting for $U_{\rm eff}$ may give important insights into more refined details of the blockade phenomena.

\section*{Acknowledgement}
R.S. and D.R. acknowledge funding from the Department of Science and Technology, India, via the Ramanujan Fellowship. D.R. also acknowledges funding from the Ministry of Electronics \& Information Technology (MeitY), India under grant for “Centre for Excellence in
Quantum Technologies” with Ref. No. 4(7)/2020-ITEA.

\appendix
\section{Complex on-site interaction for side-coupled qubits}\label{SideUeffDerivation}

To calculate the complex on-site interaction ($U_{\rm eff}$) for the side-coupled qubits medium, we match the transmission probabilities of the THLE and the QCA for a single resonator-qubit pair. To calculate the steady-state solutions of the THLE, we consider a truncated set of the operators $S^T = \left[\begin{matrix}S_{01} & \overline{S_{01}} & S_{11} & S_{12} & \overline{S_{12}} & S_{02} &\overline{S_{02}}\end{matrix}\right]$. The THLEs for these operators are given by Eqs.~\ref{QubitS01}-\ref{QubitS12} and their conjugates. In contrast to direct-coupled qubits case, we require a larger operator set to calculate $U_{\rm eff}$ in order to perfectly match the modified QCA results with those from the THLE analysis. 

After solving the set of seven coupled THLEs, we obtain the expected steady-state solution for $\mathcal{S}_{01}^{\infty}$ in terms of $\mathcal{S}_{11}^{\infty}$ as
\begin{equation}
    \mathcal{S}_{01}^{\infty} = E_1 + E_2\mathcal{S}_{11}^{\infty}, \label{b1}
\end{equation}
and the expectation value of $\mathcal{S}_{11}^{\infty}$ is given as
\begin{equation}
    \mathcal{S}_{11}^{\infty} = \frac{-\Omega_L(A^* E_1^* + AE_1)}{g_{1}(A+A^*) + \Omega_L(AE_2 + A^* E_2^*)},
\end{equation}
where,
\begin{widetext}
\begin{eqnarray}
    E_1 &=& \frac{-g_{1}\Omega_L (i\delta\omega_{q_1}AA^* + 2g_{1}^2A^* + g_{1}A + 2iUAA^*)(i\delta\omega_{q_1}A + g_{1}^2 + iUA)(i\delta\omega_{q_1}A+g_{1}^2) AA^*}{(i\delta\omega_{q_1}A+g_{1}^2) \Big((i\delta\omega_{q_1}AA^* + 2g_{1}^2A^* + g_{1}A + 2iUAA^*)(i\delta\omega_{q_1}A + g_{1}^2 + iUA)(i\delta\omega_{q_1}A+g_{1}^2)+2iUg_{1}^2\Omega_L^2A^2 \Big)},\nonumber \\ \\
    E_2 &=& \frac{4iUg_{1}\Omega_L(i\delta\omega_{q_1}A + g_{1}^2 + iUA)A^*}{(i\delta\omega_{q_1}AA^* + 2g_{1}^2A^* + g_{1}A + 2iUAA^*)(i\delta\omega_{q_1}A + g_{1}^2 + iUA)(i\delta\omega_{q_1}A+g_{1}^2)AA^* + 2iUg_{1}^2\Omega_L^2A}, \label{b4} \nonumber \\
\end{eqnarray}
\end{widetext}
where, $A = i\delta\omega_{r_1} + \Gamma$ and $\Gamma = \Gamma_L + \Gamma_R$. By integrating out the resonators we have obtained the expected value $F_{01}^{\infty}$ in Eq.~\ref{longTF01}

The quasi-classical equations for the side-coupled medium at steady-state for $N=1$ are
\begin{align}
     &(i\delta\omega_{r_1}+\Gamma)\alpha_{1}^{ss} +ig_1\beta_{1}^{ss} +i\Omega_L=0, \label{SCQ_QCA_SS_1_A} \\
     &i\delta\omega_{q_1}\beta_{1}^{ss} + 2iU|\beta_{1}^{ss}|^2\beta_{1}^{ss} + ig_1\alpha_{1}^{ss}=0. \label{SCQ_QCA_SS_1_B}
\end{align}
The transmission probabilities for the THLE and QCA for $N=1$ at steady-state is given by
\begin{align}
    \mathcal{T}_{\rm THLE} &= \frac{2\Gamma_R}{v_g I_{\rm in}}F_{11}^{\infty} \label{F11_SP}, \\
    \mathcal{T}_{\rm QCA} &= \frac{2\Gamma_R}{v_g I_{\rm in}}|\alpha_{1}^{ss}|^2,
\end{align}
where $F_{11}^{\infty}$ is given by Eq. \ref{F11final}. Similar to direct-coupled medium's case, we match the steady-state THLE and QCA transmission probabilities by demanding $|\alpha_{1}^{ss}|^2 = F_{11}^{\infty}$. Again, we choose the phase of $\alpha_{1}^{ss}$ to be the same as $F_{01}^{\infty}$ which gives $\alpha_{1}^{ss}$ as
\begin{equation} \label{alphaSCQ_ss}
    \alpha_{1}^{ss} = \frac{\sqrt{F_{11}^{\infty}} F_{01}^{\infty}}{|F_{01}^{\infty}|}.
\end{equation}
The Eq. \ref{SCQ_QCA_SS_1_A} also gives us the expression for $\beta_{1}^{ss}$ which is $\beta_{1}^{ss} = -(\Omega_L + (\delta\omega_{r_1}-i\Gamma)\alpha_{1}^{ss})/g_{1}$.
Finally, we replace $U$ with $U_{\rm eff}$ in Eq. \ref{SCQ_QCA_SS_1_B} and solve for $U_{\rm eff}$ which is given by
\begin{equation}
    U_{\rm eff} = - \frac{\delta\omega_{q_1} \beta_{1}^{ss} + g_{1} \alpha_{1}^{ss}}{2|\beta_{1}^{ss}|^2 \beta_{1}^{ss}}.
\end{equation}

The transmission probability even for a single qubit-resonator pair, given by Eqs. \ref{F11_SP}, \ref{F11final}, \ref{b1}-\ref{b4}, is a very long cumbersome expression analytically. However, several features of this system are present even at low intensity which can be captured by single-particle calculation presented in Appendix~\ref{app_rv1}.

\section{Inelastic scattering of the photons} \label{AppElasticScattering}

To investigate the elastic and inelastic components of photon transmission, we first calculate the transmission amplitude ($t_{N}(\omega_p)$) for each of the mediums of size $N$ which is given by
\begin{equation}
    t_{N}(\omega_p) = \frac{-2i\pi g_R }{E_p} \langle \chi_N \rangle,
\end{equation}
where, $\chi_N$ is $b_N$ for the direct-coupled qubits medium and $f_N$ for the side-coupled qubits medium.

\begin{figure} 
    \centering
    \includegraphics[width= \linewidth]{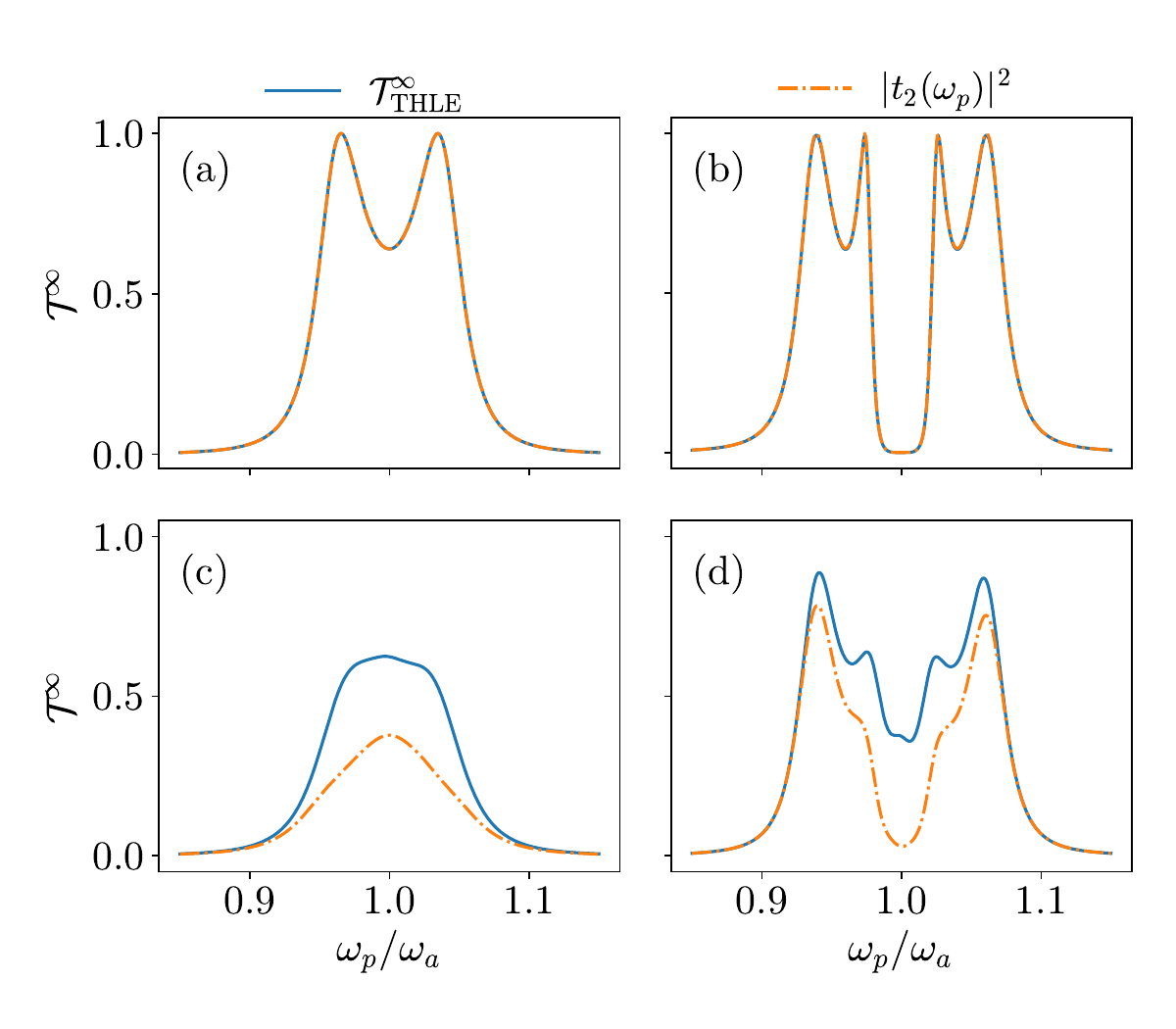}
    \caption{Transmission profiles comparing the mod-square of transmission amplitude~($|t_{2}(\omega_p)|^2$) with transmission probabilities~($\mathcal{T}^{\infty}_{\rm THLE}$) for two ($N=2$) direct-coupled qubits $(a,c)$ and two side-coupled qubits $(b,d)$. The incident photon intensity $I_{\rm in}$ (in terms of $\omega_a / v_g$) increases on going down the column as $I_{\rm in}=1.0 \times 10^{-6} (a,b)$ and $I_{\rm in}=0.011 (c,d)$. Parameters are: $\omega_{q_j}=\omega_{r_j}=\omega_a$, $J_x = \Gamma_L=\Gamma_R=0.02 \omega_a$, $U=1.05 \omega_a$, $g_j=0.04\omega_a$. }
    \label{app1}
\end{figure}
Interestingly, the mod-squared transmission amplitude $|t_{N}(\omega_p)|^2$ does not give us the total transmission probability $(\mathcal{T}_{\rm THLE})$. It only captures what is called coherent component of transmission \cite{Hoi2011} as opposed to an incoherent component that displays inelastic scattering of the photons at higher intensity. To investigate the coherent and incoherent contribution in the transmission probability, we compare the mod-squared transmission amplitudes ($|t_{N}(\omega_p)|^2$) with the transmission probability ($\mathcal{T}^{\infty}_{\rm THLE}$), at $m=6$, for the two mediums of size $N=2$ in Fig. \ref{app1}. At low light intensities, we find that $|t_{2}(\omega_p)|^2$ is equal to $\mathcal{T}^{\infty}_{\rm THLE}$ in Fig. \ref{app1}(a,b), which indicates that there is very little incoherent contribution to the transmission due to the nonlinearities at such intensities. As we increase the intensity, we find that the difference between $\mathcal{T}^{\infty}_{\rm THLE}$ and $|t_{2}(\omega_p)|^2$ increases, and the coherent part $|t_{2}(\omega_p)|^2$ is less than the total transmission in Fig. \ref{app1} (c,d). The difference between  $\mathcal{T}^{\infty}_{\rm THLE}$ and $|t_{2}(\omega_p)|^2$ provides an estimate of incoherently transmitted photons in our mediums.

\begin{figure}
    \centering
    \includegraphics[width= \linewidth]{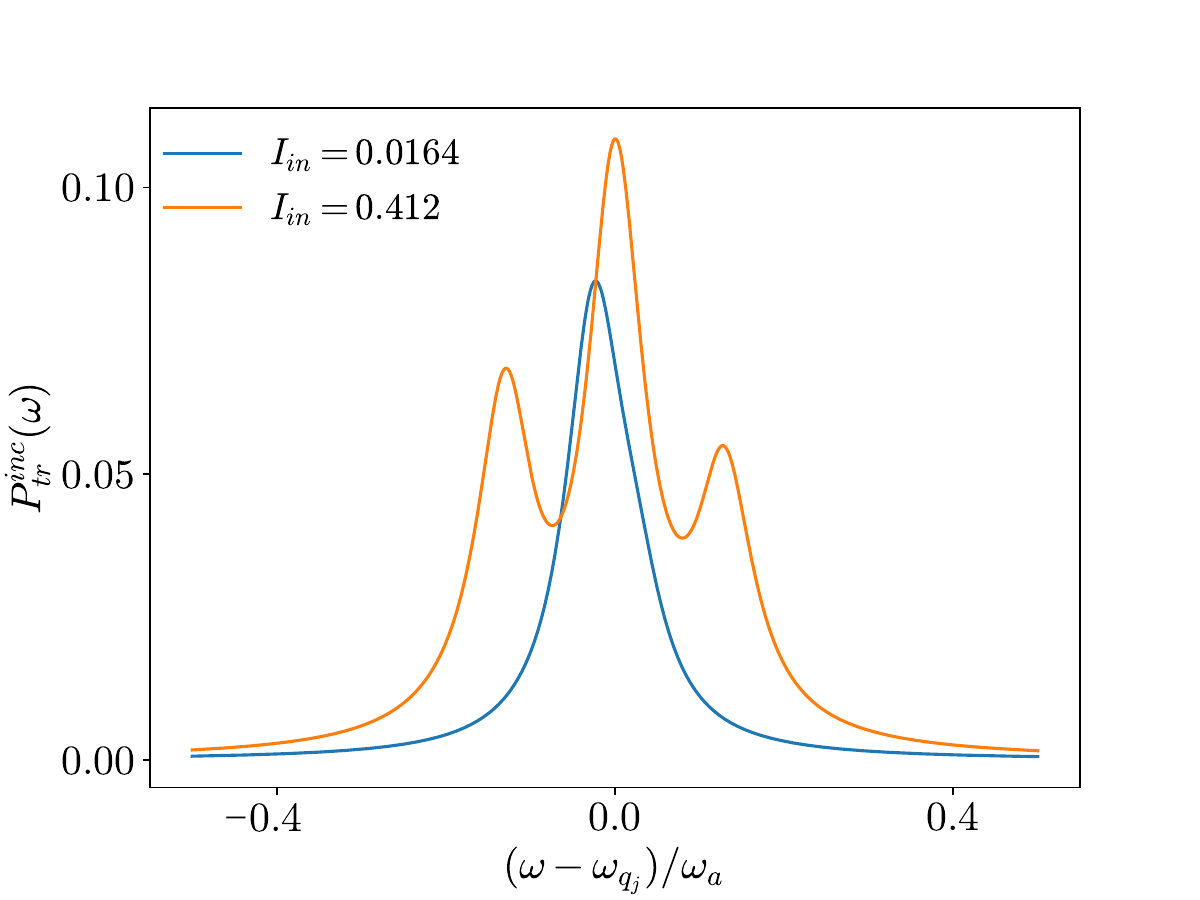}
    \caption{The incoherent component of power spectrum showing inelastic scattering of photons (the two side peaks at higher intensity) for two direct-coupled qubits at resonant input frequency. Intensity ($I_{\rm in}$) is in terms of $\omega_a / v_g$. Parameters are: $\omega_{q_j}=1.0\omega_a, J_x = \Gamma_L=\Gamma_R=0.02 \omega_a$, $U=1.05 \omega_a$.}
    \label{app3}
\end{figure}
As only lossless systems have been considered in the paper, the presence of inelastic components can be confirmed by evaluating the power spectrum of the transmitted photons. In the direct-coupled qubits, it can be obtained by \cite{Roy2017}:
\begin{equation}
    P_{tr}(\omega) = Re \int^{\infty}_{0} \frac{d\tau}{\pi} e^{i \omega \tau} \langle a_{R_k}^{\dagger}(t)a_{R_k}(t+\tau) \rangle.
\end{equation}
Under the initial condition as mentioned in the main text, the above expression reduces down to 
\begin{equation}
    P_{tr}(\omega) = \frac{2 \Gamma_R}{\pi v_g}Re \int^{\infty}_{0} d\tau e^{i \omega \tau} \langle b^{\dagger}_{N}(t)b_{N}(t+\tau) \rangle.
\end{equation}
The procedure to calculate the two-time correlation of qubit operators $\langle b^{\dagger}_{N}(t)b_{N}(t+\tau) \rangle$ for a single qubit is mentioned in \cite{Roy2017}. The procedure can be extended to multiple direct-coupled qubits to evaluate the transmission power spectrum having coherent and incoherent components, i.e, $P_{tr}(\omega)=P^{coh}_{tr}(\omega)+P^{inc}_{tr}(\omega)$. For a single qubit, the coherent component is $P^{coh}_{tr}(\omega)=(2\Gamma_R /v_g)| \mathcal{S}_{01}(t \rightarrow \infty)|^2 \delta(\omega-\omega_p)$ while the incoherent component is $P^{inc}_{tr}(\omega)=(\Gamma_R/\pi) Re(I_3)$ where $I_3$ is given as
\begin{widetext}
\begin{equation}
I_3 = \frac{|\mathcal{S}_{01}(\infty)|^2 - \mathcal{S}_{11}(\infty)}{\mu_1} + \frac{\Omega_L^2(\mathcal{S}_{10}(\infty))^2-\Omega_L^2 (|\mathcal{S}_{01}(\infty)|^2-\mathcal{S}_{11}(\infty))\mu_2 / \mu_1 - 2i\Omega_L\mu_2\mathcal{S}_{10}(\infty)\mathcal{S}_{11}(\infty)}{2 \mu_1 \mu_2 \mu_3 + \Omega_L^2 (\mu_1+\mu_2)},
\end{equation}
\end{widetext}
where $\mu_1=-(\Gamma_L+\Gamma_R) + i(\omega-\omega_{q_1})$, $\mu_2=-(\Gamma_L+\Gamma_R) + i(\omega+\omega_{q_1}-2\omega_p)$ and $\mu_3=-(\Gamma_L+\Gamma_R) + i(\omega-\omega_{p})$. The above expressions are also valid for larger system sizes and their power spectrum can be obtained by replacing the operators with the ones for the last qubit. 

The incoherent part of transmission power spectrum for two direct-coupled qubits is shown in Fig.~\ref{app3}. At low incident intensity, we observe a single peak around the qubits' frequency. However, as the incident intensity is increased, we observe three peaks. The three peaks at high input light intensity are known as the Mollow triplets. These triplets are characteristic of the inelastically scattered photons \cite{Astafiev10}.

\begin{figure}
    \centering
    \includegraphics[width= \linewidth]{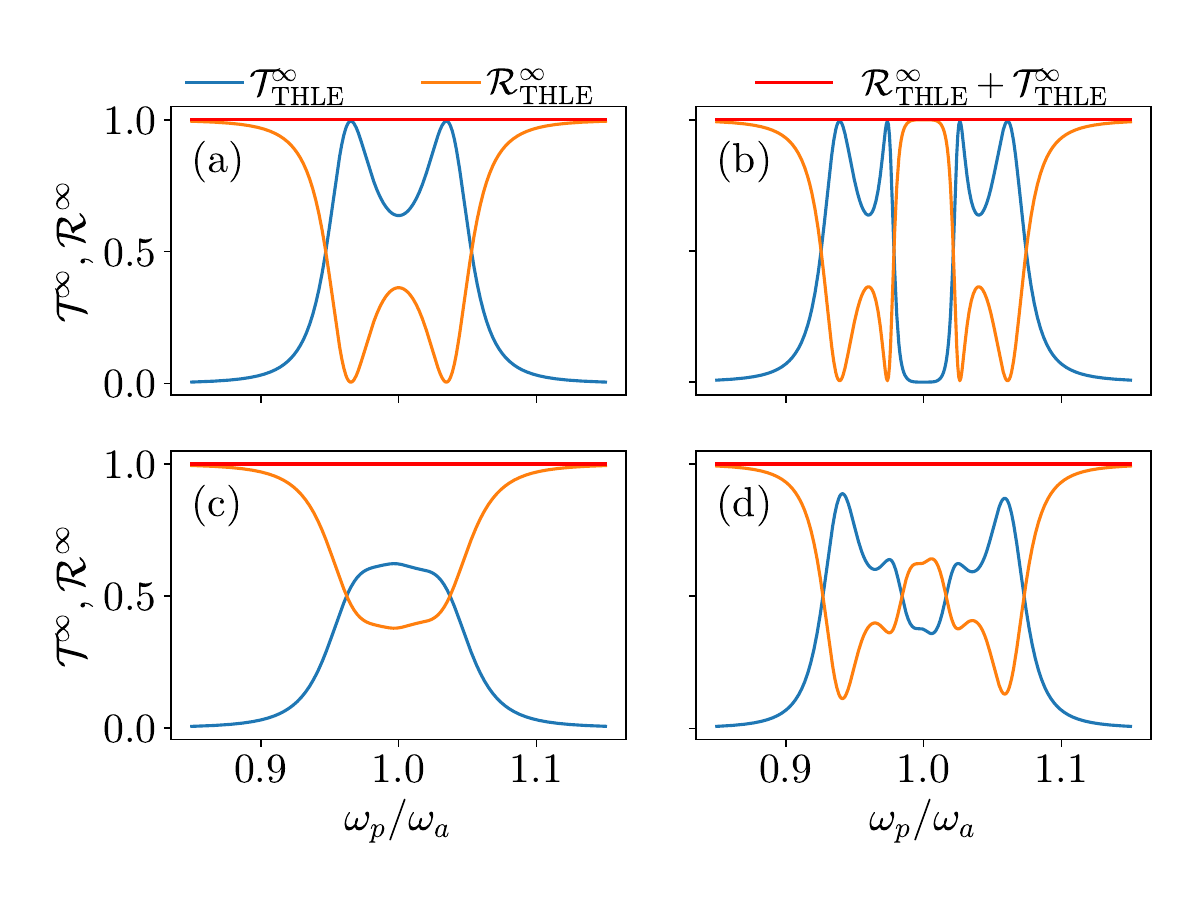}
    \caption{The transmission and reflection  probabilities along with their sum with increasing $\omega_p$ for a system of two direct-coupled qubits $(a,c)$ and two side-coupled qubits $(b,d)$. Photon intensity $I_{\rm in}$ (in terms of $\omega_a / v_g$) increases on going down the column as $I_{\rm in}=1.0 \times 10^{-6} (a,b)$ and $I_{\rm in}=0.011 (c,d)$. Parameters are: $\omega_{q_j}=\omega_{r_j}=\omega_a$, $J_x = \Gamma_L=\Gamma_R=0.02 \omega_a$, $U=1.05 \omega_a$, $g_j=0.04\omega_a$.}
    \label{app2}
\end{figure}
Further, to demonstrate that the mediums considered here are lossless, we numerically verify that the sum of steady-state reflection and transmission probabilities is unity in Fig.~\ref{app2}. This is true even at higher intensities, Fig.~\ref{app2} (c,d), where the effects of interactions are active.

\section{Single-photon transport in a side-coupled medium} \label{app_rv1}

To understand the single-photon transport dynamics in a single resonator-qubit pair, we first write the side-coupled qubits system's Hamiltonian given in Section~\ref{SCQ_Model} in real-space in which the momentum space operators are replaced by their Fourier transform. The real-space Hamiltonian is thus given as
\begin{align}
    \frac{H_r}{\hbar} = &\frac{H_{side}}{\hbar} - i v_g \int dx \Big( a^{\dagger}_{Lx} \frac{\partial}{\partial x} a_{Lx} + a^{\dagger}_{Rx} \frac{\partial}{\partial x} a_{Rx} \Big) \nonumber \\
    &+(\sqrt{2v_g \Gamma_L}f^{\dagger}_1 a_{L0} + \sqrt{2v_g \Gamma_R}f^{\dagger}_1 a_{R0} +H.c.).
\end{align}
Similar to the main text, we assume a photon is incoming from the left of the medium. For the side-coupled medium of a single qubit-resonator pair in the ground state, the wave function of an incoming single photon is given by
\begin{equation}
    |\psi \rangle_{in} = \frac{1}{\sqrt{2\pi}} \int dx e^{ikx} a^{\dagger}_{Lx} |\phi \rangle \otimes |0\rangle_r \otimes |0\rangle_q,
\end{equation}
where $k$ is the wave vector of the incoming photon and $|\phi \rangle$ denotes the vacuum of light field. $|0\rangle_r$ and $|0\rangle_q$ indicate the ground state of the resonator and qubit, respectively. An incoming single photon can excite the resonator and qubit to their excited state $|1\rangle_r$ and $|1\rangle_q$. Therefore, the wave function of the system after the photon interacts with the medium ($|\psi \rangle_{s}$) is given by
\begin{align}
    |\psi \rangle_{s} &= \frac{1}{\sqrt{2\pi}} \int dx \{\delta(x)[e_{k}^{r}|\phi \rangle \otimes |1\rangle_r \otimes |0\rangle_q \nonumber \\
    &+ e_{k}^{q}|\phi \rangle \otimes |0\rangle_r \otimes |1\rangle_q]+ [\Phi^{L}_{k} (x)a_{Lx}^{\dagger} \nonumber \\
    &+ \Phi^{R}_k (x)a_{Rx}^{\dagger}]|\phi \rangle \otimes |0\rangle_r \otimes |0\rangle_q
    \},
\end{align}
where $e^{q}_{k}$ and $e^{r}_{k}$ are the amplitudes of the excited resonator and qubit, respectively. Here, $\Phi^L_k(x)$ and $\Phi^R_k(x)$ are the amplitudes of the scattered photon in the left and right sides of the side-coupled medium.

These amplitudes can be found by the time-independent Schr\"{o}dinger equation $H_r |\psi \rangle_s = \hbar \omega_p |\psi \rangle_s $, which gives two coupled linear equations for the amplitudes of the resonator and the qubit, and two more inhomogeneous differential equations for the amplitudes of photons. The two differential equations are
\begin{align}
    &-iv_g \frac{\partial}{\partial x}\Phi^{L}_{k}(x) + \sqrt{2v_g \Gamma_L}e_k^{r}\delta(x) = \omega_p \Phi^{L}_{k}(x), \label{SP_bath_L} \\
    &-iv_g \frac{\partial}{\partial x}\Phi^{R}_{k}(x) + \sqrt{2v_g \Gamma_R}e_k^{r}\delta(x) = \omega_p \Phi^{R}_{k}(x). \label{SP_bath_R}
\end{align}
By following \cite{Manasi2018}, the Eqs.~\ref{SP_bath_L} and \ref{SP_bath_R} are regularized (at $x=0$) and we assume photon coming from the left of the medium which gives
\begin{align}
    \Phi^{L}_{k}(0) &= 1 - i\sqrt{\frac{\Gamma_L}{2 v_g}} e^{r}_{k}, \label{SP_PhiL0}\\
    \Phi^{R}_{k}(0) &= - i\sqrt{\frac{\Gamma_R}{2 v_g}} e^{r}_{k}. \label{SP_PhiR0}
\end{align}
The initial states eq. \ref{SP_PhiL0} and \ref{SP_PhiR0} are then substituted in the solutions of the medium elements. Therefore, we are left with the equations of resonator and qubit amplitudes which are written as
\begin{align} 
     &(\omega_{r_1}-\omega_p - i(\Gamma_L+\Gamma_R))e^{r}_{k} + g_{1}e^{q}_{k} = - \sqrt{2v_g \Gamma_L}, \label{diffeqn1} \\
     &(\omega_{q_1}-\omega_p)e^{q}_{k} + g_{1}e^{r}_{k}=0. \label{diffeqn2}
\end{align}

 The transmission amplitude for this medium is given by
\begin{equation} \label{tN}
    t^{s}_{1}(\omega_p) = -i\sqrt{\frac{2\Gamma_R}{v_g}} e^{r}_{k}.
\end{equation}
By using the solutions of Eq.~\ref{diffeqn1} and \ref{diffeqn2}, we obtain the transmission amplitude for a single-site side-coupled qubit medium $t^s_1(\omega_p)$ as
\begin{equation} \label{ts1}
    t^s_1(\omega_p) = \frac{2i\sqrt{\Gamma_L \Gamma_R}(\omega_{q_1} - \omega_p)}{(\omega_{r_1} - \omega_p - i(\Gamma_L+\Gamma_R))(\omega_{q_1} - \omega_p) - g_{1}^2}.
\end{equation}
The numerator of Eq.~\ref{ts1} explains the complete reduction of transmission at resonance with the qubit. The mod-squared transmission amplitude for the qubit-resonator pair $|t^s_1(\omega_p)|^2$ shows two peaks at the photon frequencies
\begin{equation}
\omega_{p_\pm} = \Big( \frac{\omega_{r_1} + \omega_{q_1}}{2} \Big) \pm \sqrt{\Big( \frac{\omega_{r_1} - \omega_{q_1}}{2} \Big)^2 + g_{1}^2}.
\end{equation}

\section{The self-consistent mean-field analysis for the direct-coupled system} \label{SCMF}
The propagation of photons through the medium can also be studied using the mean-field approximation applied on the on-site interaction term in the Hamiltonian. In the case of direct-coupled qubits, after applying the mean-field approximation the Hamiltonian of the the medium becomes
\begin{align}
    H^{mf}_{direct} = &\hbar\sum_{j=1}^{N}\omega_{q_j}b_j^{\dagger}b_j + Ub_j^{\dagger}b_j(\langle b_j^{\dagger}b_j \rangle - 1)
     \nonumber \\ \quad &+ \hbar\sum_{j=1}^{N-1}2J_x(b_j^{\dagger}b_{j+1} + b_{j+1}^{\dagger}b_j).
\end{align}
The bath and bath coupling terms are also included in the complete Hamiltonian of the medium. 
To obtain the mean-field equations for the system, we derive a set of $N$ equations by using the Heisenberg equations for the operators $b_j$ for each site given by
\begin{align} \label{MF_direct}
    \dot{b_j} = &-(i\delta\omega_{q_j} + \Gamma_j)b_j - iU(\langle  b^{\dagger}_jb_j \rangle-1)b_j \nonumber \\
    &- 2iJ_x(b_{j-1} + b_{j+1}) -i\Omega_L\delta_{1,j},
\end{align}
where, $\Gamma_j=0$ for $2 \leq j \leq N-1$, $\Gamma_1=\Gamma_L$ and $\Gamma_N=\Gamma_R$. If $N=1$, $\Gamma = \Gamma_L+\Gamma_R$. We use open boundary condition for the medium, i.e. $b_0=b_{N+1}=0$. Similar to the quasi-classical equations, the mean-field equations has no dependence on other excited states and, therefore, $N$ mean-field equations have all the necessary information to model the medium.

By comparing the Eq.~\ref{MF_direct} with the quasi-classical equations, we observe that the on-site interaction term in mean-field equations has missed a factor of 2 and has an additional -1 as well. Therefore, we put the factor of 2 in the on-site interaction and remove -1 explicitly in the numerical calculations.

Each of the mean-field equations has number operator~($\langle b_j^{\dagger}b_j  \rangle$) in the RHS. Therefore to solve these equations at steady-state, we multiply each of the $N$ equations with its conjugate to derive a set of coupled equations for number operators at each site. For example, we get the following self-consistent mean-field equations for a medium two direct-coupled qubits 
\begin{widetext}
\begin{align}
    \langle b_1^{\dagger} b_1 \rangle &= \frac{| \delta\omega_{q_2} + 2U\langle b_2^{\dagger} b_2 \rangle  - i\Gamma_R |^2 \Omega_L^2}{| 4J_x^2 - (\delta\omega_{q_1} + 2U \langle b_1^{\dagger} b_1 \rangle  - i\Gamma_L )(\delta\omega_{q_2} + 2U\langle b_2^{\dagger} b_2 \rangle  - i\Gamma_R ) |^2}, \\
    \langle b_2^{\dagger} b_2 \rangle &= \frac{4J_x^2 \Omega_L^2}{| 4J_x^2 - (\delta\omega_{q_1} + 2U\langle b_1^{\dagger} b_1 \rangle  - i\Gamma_L )(\delta\omega_{q_2} + 2U\langle b_2^{\dagger} b_2 \rangle  - i\Gamma_R ) |^2}.
\end{align}
\end{widetext}

In order to get numerical solutions of these equations we run a self-consistent loop over the set of equations for a large number of iterations and calculate the output transmission using the formula
\begin{equation}
    \mathcal{T}_{mf}^{\infty} = \frac{2\Gamma_R}{v_g I_{\rm in}} \langle b_N^{\dagger}b_N \rangle.
\end{equation}

As mentioned in the main text, the self-consistent mean-field after making the corrections mentioned above agrees with QCA for a wide range of parameters.

\bibliography{References}
\end{document}